\documentclass[aps,prb,twocolumn,floatfix,showpacs,superscriptaddress]{revtex4-2}
\usepackage{graphics}
\usepackage{epsfig}
\usepackage{times}
\usepackage{bm}
\usepackage{braket}
\usepackage{color}
\usepackage{multirow}
\usepackage{dcolumn}

\usepackage{amsmath}	
\begin{document}
\title{Comparing quantum fluctuations in the spin-$\frac{1}{2}$ and spin-$1$ XXZ Heisenberg models\\ on square and honeycomb lattices}
\author{Masahiro Kadosawa}
\affiliation{Department of Physics, Chiba University, Chiba 263-8522, Japan}
\author{Masaaki Nakamura}
\affiliation{Department of Physics, Ehime University, Ehime 790-8577, Japan}
\author{Yukinori Ohta}
\affiliation{Department of Physics, Chiba University, Chiba 263-8522, Japan}
\author{Satoshi Nishimoto}
\affiliation{Department of Physics, Technical University Dresden, 01069 Dresden, Germany}
\affiliation{Institute for Theoretical Solid State Physics, IFW Dresden, 01069 Dresden, Germany}

\date{\today}

\begin{abstract}
We present a detailed investigation of the XXZ Heisenberg model for spin-$1/2$ and spin-$1$ systems on square and honeycomb lattices. Utilizing the density-matrix renormalization group (DMRG) method, complemented by Spiral Boundary Conditions (SBC) for mapping two-dimensional (2D) clusters onto one-dimensional (1D) chains, we meticulously explore the evolution of staggered magnetization and spin gaps across a broad spectrum of easy-axis anisotropies. Our study reveals that, despite the lower site coordination number of honeycomb lattice, which intuitively suggests increased quantum fluctuations in its N\'eel phase compared to the square lattice, the staggered magnetization in the honeycomb structure exhibits only a marginal reduction. Furthermore, our analysis demonstrates that the dependence of staggered magnetization on the XXZ anisotropy $\Delta$, except in close proximity to $\Delta=1$, aligns with series expansion predictions up to the 12th order. Notably, for the $S=1/2$ honeycomb lattice, deviations from the 10th order series expansion predictions near the isotropic Heisenberg limit emphasize the critical influence of quantum fluctuations on the spin excitation in its N\'eel state. Additionally, our findings are numerically consistent with the singular behavior of the spin gap near the isotropic Heisenberg limit as forecasted by spin-wave theory. The successful implementation of SBC marks a methodological advancement, streamlining the computational complexity involved in analyzing 2D models and paving the way for more precise determinations of physical properties in complex lattice systems.
\end{abstract}

\maketitle

\section{introduction}

In the complex world of quantum magnetism, the interplay between spin interactions and lattice geometry crafts a fascinating landscape of ground states and excitations (e.g., see Refs.~[\onlinecite{schollwoeck2004,lacroix2011,sachdev2011}]). A central subject for this exploration is the XXZ Heisenberg model~\cite{Heisenberg1928},  a cornerstone that has deeply enriched our comprehension of anisotropic magnetic systems~\cite{Lines1963,Achiwa1969}. The model, celebrated for its versatility in representing real materials, allows for the examination of quantum fluctuations -- the quintessential quantum mechanical effect that destabilizes classical magnetic order, paving the way for the emergence of novel quantum phases like spin liquids~\cite{savary2016,zhou2017}.

This paper focuses on comparing the manifestations of quantum fluctuations within the
$S=1/2$ and $S=1$ XXZ models on two fundamentally distinct lattice structures: the square and the honeycomb. These lattices, emblematic of different coordination environments and geometric constraints, provide a compelling backdrop against which the interplay of spin magnitude and lattice topology can be meticulously dissected. The square lattice, with its direct links to high-temperature superconductivity~\cite{schrieffer2007} and magnetic order~\cite{sachdev2011} in solid-state compounds, and the honeycomb lattice, notable for hosting exotic phenomena such as the quantum spin Hall effect~\cite{bernevig2006} and potential quantum spin liquid states~\cite{Yi2017}, are ideal platforms for this comparative study.

Numerical simulations of such systems often pose significant challenges. When instantiated on a finite-size lattice, the total degrees of freedom exponentially increase with lattice size. This constraint on the geometry and size of the cluster becomes particularly notable for systems in more than two dimensions. Consequently, an extrapolation of physical quantity to an infinite system size is imperative to ascertain the bulk value. However, the execution of such finite-size scaling is usually fraught with challenges due to the presence of multiple scaling dimensions, such as the x and y directions in a 2D case. In our previous studies~\cite{Kadosawa2022,Kadosawa2023-1}, we introduced an efficient numerical method for determining the local order parameter in 2D systems through the use of spiral boundary conditions (SBC). This method provides a promising approach to address the challenges associated with finite-size scaling in numerical simulations.

Applying SBC allows for the exact projection of lattice models, even those extending beyond 2D, onto 1D periodic chains that maintain translational symmetry. Within this projected 1D chain, each lattice site is denoted by a single coordinate, contrary to the dual coordinates used in the original 2D cluster. This simplification means that only one finite-size scaling analysis along the chain direction is necessary to ascertain a physical quantity in the thermodynamic limit. We have demonstrated the capability of precisely determining the magnitude of staggered magnetization for the XXZ Heisenberg model on a square lattice ranging from $S=1/2$ to $6$~\cite{Kadosawa2023-1}. We here consider the extension of this technique to studies of the honeycomb-lattice model and further demonstrates the systematic calculation of excitation energy in the bulk limit.

In this paper, we investigate the $S=1/2$ and $S=1$ XXZ Heisenberg models on
square and honeycomb lattices employing the DMRG method. We delve into
the evolution of staggered magnetization and the accompanying spin gap with
easy-axis anisotropy. It is shown that by applying SBC to both lattice types,
a finite-size scaling analysis towards the thermodynamic limit can be effortlessly
conducted for the studied physical quantities. The efficacy of our approach is
corroborated by comparing our findings with pre-existing numerical and
analytical results. Our analysis reveals that for most cases, the staggered
magnetization and spin gap within the easy-axis N\'eel phase can be approximately
accounted for by series expansions (SE) of the 10th to 12th order in terms of
$1/\Delta$. However, for the  $S=1/2$ honeycomb lattice, the results significantly
deviate from those of the 10th order SE across a broad range near the isotropic
Heisenberg limit due to strong quantum fluctuations. Furthermore, for all models
considered, we obtain results that are numerically consistent with the singular
behavior of the spin gap near the isotropic Heisenberg limit as predicted by
spin-wave theory (SWT). Also, we find a marked reduction in quantum fluctuations
transitioning from $S=1/2$ to $S=1$ across all physical quantities assessed.

The paper is structured as follows: Sec. II provides a detailed description of our spin model. In Sec. III, we elucidate the method of mapping 2D models to 1D using SBC, along with the procedures for calculating physical quantities via the DMRG technique. Sec. IV presents our numerical findings, examining the influence of lattice type, XXZ anisotropy magnitude, and spin size on the stability of staggered magnetization and the magnitude of the spin gap. Additionally, we incorporate a discussion on the specific behavior of the spin gap for the $S=1/2$ honeycomb-lattice case. Finally, in Sec. V, we conclude the paper with a summary and further insights into the observed phenomena.

\section{model}

The Hamiltonian of the XXZ Heisenberg model is represented as follows:
\begin{align}
\mathcal{H} = \sum_{\braket{ij}}(S^{x}_{i}S^{x}_{j} + S^{y}_{i}S^{y}_{j} + \Delta S^{z}_{i}S^{z}_{j})
\end{align}
where $S^\gamma_i$ $(\gamma = x,y,z)$ are the spin-$S$ operators,
$\Delta$ is the anisotropy parameter, and the sum $\braket{ij}$
runs over all nearest-neighbor pairs. In this context, we consider
two types of lattice structures: the square lattice and the honeycomb
lattice. The XXZ models associated with these lattices have been
significantly examined thus far.

There are three phases depending on $\Delta$~\cite{Kubo1988,Weihong1991-2,Viswanath1994,Yunoki2002,Braiorr-Orrs2019}:
(i) For $\Delta > 1$ easy-axis N\'eel phase with antiferromagnetic
(AFM) spin alignment along the $z$-direction,
(ii) for $-1 < \Delta < 1$ easy-plane N\'eel (XY) phase with
AFM spin alignment along some arbitrary direction in the
$xy$-plane, and (iii) for $\Delta < -1$ ferromagnetic (FM)
phase with fully-polarized spins along the $z$-direction.
The phase transitions at $\Delta=\pm1$ are both first order.
For $\Delta=-1$ this model can be exactly solved:
the N\'eel and FM states are degenerate at the ground
state where the energies are $E_0=-2NS^2$ and
$E_0=-(3/2)NS^2$ for square- and honeycomb-lattice models.
At $\Delta=-1^+$, the wave function of N\'eel state is
expressed as
\begin{align}
|\Psi_0({\rm XY})\rangle=\sum_m \lambda_m |\psi_m \rangle,
\label{eq:XYstate}
\end{align}
where $|\psi_m \rangle$ are bases restricted to
$S^z_{\rm tot}=\sum_{i=1}^L \langle S^z_i \rangle=0$
subspace, $m$ is summed over all possible combinations of the
spin configurations, and $\lambda_m$ are determined for
each $S$~\cite{Kadosawa2023-1}. The magnitude of staggered
magnetization is $S$ with the direction parallel to the
$xy$-plane. At $\Delta=-1^-$, the function of FM state is
\begin{align}
|\Psi_0({\rm FM})\rangle=\frac{1}{\sqrt{2}}(|\Uparrow \rangle+|\Downarrow \rangle),
\label{eq:FMstate}
\end{align}
where $|\Uparrow \rangle$ and $|\Downarrow \rangle$ denote
fully-polarized states toward $z$ and $-z$ directions,
respectively. These states $|\Psi_0({\rm XY})\rangle$ and
$|\Psi_0({\rm FM})\rangle$ are orthogonal.

In regards to the square lattice model, it has been numerically confirmed that for
$S=1/2$~\cite{Bishop2016,Kadosawa2023-1}, N\'eel long-range order (LRO) always exists
for $\Delta>-1$. On the other hand, for the honeycomb-lattice model, due to fewer bonds
between adjacent sites compared to the square lattice, quantum fluctuations are larger,
and there is not yet a complete consensus on which $S$ and $\Delta$ regions stabilize
N\'eel LRO~\cite{Affleck1988,Wojtkiewicz2023}.

\section{method}

\begin{figure}[tbh]
	\centering
	\includegraphics[width=1.0\linewidth]{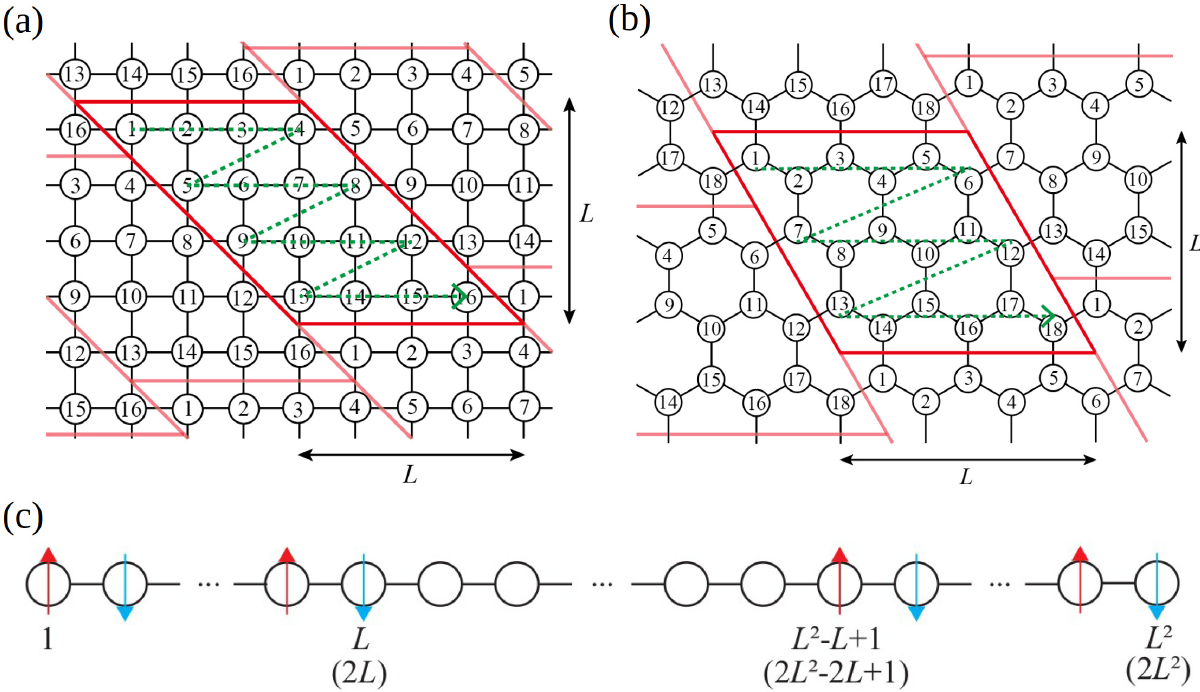}
	\caption{
		(a) 2D square-lattice $4 \times 4$ cluster and (b) 2D honeycomb-lattice $3 \times 3$ cluster. The regions outlined by red lines represent the original clusters. Through the application of SBC, the 2D clusters are projected onto 1D periodic chains, where the sites are aligned along the green lines.
		(c) Schematic depiction of the 1D open chain utilized for our DMRG calculations, focusing on staggered spin moment $m^z_\alpha(S,\Delta)$ and spin gap $\varepsilon_\alpha(S,\Delta)$. In the square lattice, pinning is applied to $L$ sites at each end, while in the honeycomb lattice, pinning is applied to $2L$ sites. Upward and downward arrows signify pinned spins with $S^z=S$ and $S^z=-S$, respectively.
	}
	\label{fig:lattice}
\end{figure}

\subsection{spiral boundary conditions}

Applying the DMRG method to 2D systems introduces significant challenges,
primarily due to two factors. Firstly, the entanglement entropy, which quantifies
the quantum correlations within different parts of the system, follows an ``area law''.
This law indicates that entanglement entropy scales with the boundary area of a region,
complicating the simulation of large systems. Secondly, the DMRG's sweeping process,
which optimizes quantum states site by site in a linear fashion in 1D systems, encounters
difficulties in the more complex geometries of 2D systems. This complexity can lead to
inaccuracies because a straightforward sweeping mechanism is harder to implement
across 2D lattices.

To address these challenges, careful management of boundary conditions is
essential for accurate DMRG simulations. Traditional approaches often employ
cylinder or torus configurations for 2D systems. Yet, these configurations can
create short bond loops and impose an unnatural periodicity on the wave function,
leading to inaccuracies such as an unexpected plaquette constraint on particles
or spins. An inappropriate choice of boundary conditions might also skew the
energy states observed in finite clusters away from those relevant in the
thermodynamic limit, instead of systematic errors due to the finite-size effects.

A promising alternative that circumvents these limitations involves the implementation
of SBC~\cite{Nakamura2021,Kadosawa2022}. SBC enables the exact projection of lattice
models, including those extending beyond two dimensions, onto 1D periodic chains that
preserve translational symmetry. This projection effectively transforms a 2D $L \times L$
cluster into a 1D chain, maintaining nearest-neighbor and ($L-1$)th-neighbor bonds for
square lattices, and nearest- and ($2L-1$)th-neighbor bonds for honeycomb lattices, as
depicted in Fig.~\ref{fig:lattice}(a,b). This innovative approach prevents the emergence of
artificial short bond loops and ensures an even distribution of quantum entanglement
across the chain, leveraging translational symmetry. 

Notably, SBC minimizes the distance of the longest bonds, denoted as $d$, to $L-1$
for square and $2L-1$ for honeycomb lattices, optimizing conditions for DMRG calculations.
In contrast, conventional periodic boundary conditions would increase $d$ to $2L$ and
$4L-2$, respectively, posing challenges for DMRG analysis.

Furthermore, SBC provides a significant benefit for finite-size scaling analysis.
By projecting the original 2D lattice onto a 1D chain, SBC enables the indexing
of each lattice site with a singular coordinate, rather than the dual-coordinate
system inherent to 2D clusters. This transition to a single-coordinate framework
simplifies the analytical process to a unidimensional scale. It facilitates a more
direct method for extrapolating physical quantities to the thermodynamic limit,
enhancing the accuracy and efficiency of our simulations.

\subsection{density-matrix renormalization group}

The investigation of the ground state of the 1D chain, transformed via SBC,
is conducted using the DMRG method~\cite{White1992}. For this purpose,
we implement open boundary conditions on the 1D chain, a choice that
significantly enhances the precision of our DMRG calculations. Our study
encompasses open chains with lengths up to $N = L^2 = 196$ sites for
the square lattice and up to $N = 2L^2 = 162$ sites for the honeycomb lattice.
To ensure the robustness of our calculations, we retain up to $m=8000$
density-matrix eigenstates, with all calculated values subsequently extrapolated
to the limit of $m\rightarrow\infty$. The maximum discarded weight observed
is on the order of $10^{-6}$.

Furthermore, we intentionally break the spin-rotation symmetry by employing
a spin pinning technique. This approach effectively lifts the degeneracy of the
ground state, thereby efficiently reducing the dimensionality of the Hilbert space
required for our calculations. As a result, even for computations of 2D systems,
the discarded weight remains minimal, enhancing the accuracy and feasibility of
our analysis.

\begin{figure}[t]
	\centering
	\includegraphics[width=1.0\linewidth]{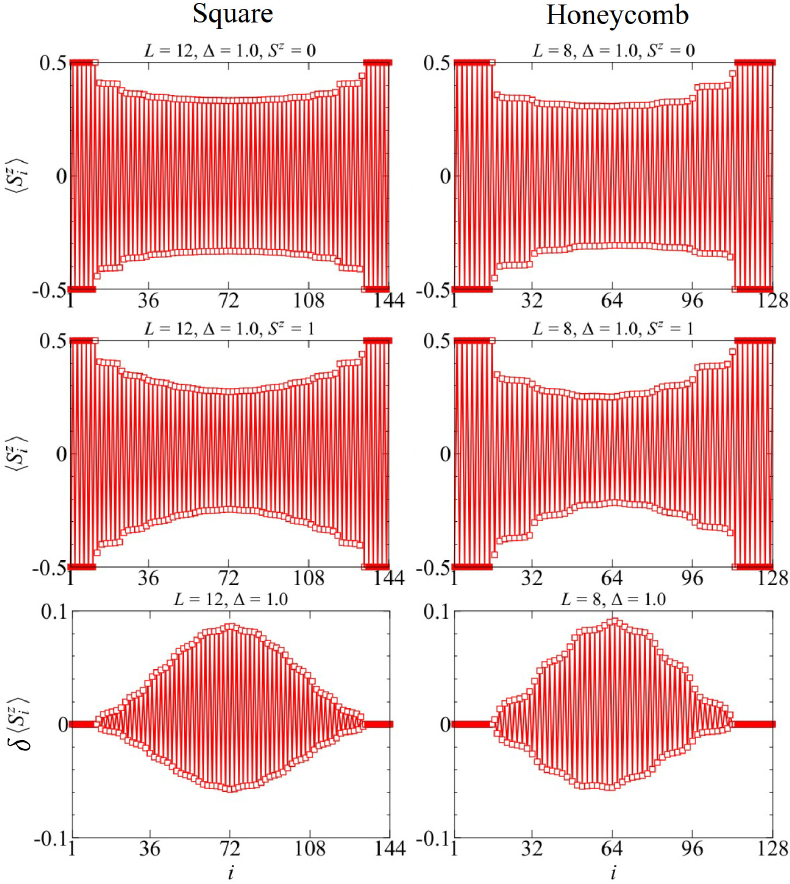}
	\caption{
		Comparative profiles of the local spin moment for $12\times12$ square
		(left panel) and $8\times8$ honeycomb (right panel) lattices at spin-isotropic
		case $\Delta=1$. The profiles depict $\braket{S^z_i}$, the expectation value
		of the $z$-component of spin at site $i$, for two distinct total spin sectors:
		$S^z=0$ (top row) and $S^z=1$ (middle row). The bottom row illustrates
		the differential profiles, showing the variance in $\braket{S^z_i}$ between
		the $S^z=0$ and $S^z=1$ sectors.
	}
	\label{fig:localSz}
\end{figure}

\subsection{physical quantities}

\subsubsection{staggered magnetization}\label{subsec:stagmag}

We examine the N\'eel state for $\Delta \ge 1$, where the magnitude
of staggered magnetization serves as the order parameter in both square
and honeycomb lattice XXZ models. In the original 2D clusters, N\'eel
order characterized by $\boldsymbol{k}=(\pi,\pi)$ is transformed into
a magnetic order with $k=\pi$ along the projected 1D chain through
our implementation of SBC. For the analysis, we utilize open chains and
determine the magnitude of staggered magnetization by measuring half
the amplitude of the Friedel oscillation of $\braket{S^z_i}$ from the system
edges. Given the condition $\Delta \ge 1$, it is sufficient to focus on
the $z$-component of the spin moment.

To establish such an open chain configuration, we sever $L$ ($L+1$) bonds
between adjacent sites on the projected 1D periodic chain for square
(honeycomb) lattices, as described in recent
works~\cite{Kadosawa2023-1,Kadosawa2023-2}. We specifically examine the
local moments of the two central spins, $\braket{S^{z}_{N/2}}$ and
$\braket{S^{z}_{N/2+1}}$, employing spin pinning near the system edges,
such as setting $\braket{S^z_i}=(-1)^{i-1} S$ at sites $i=1,\dots,L$ and
$i=L^2-(L-1),\dots,L^2$ for the square lattice; at sites $i=1,\dots,2L$ and
$i=2L^2-(2L-1),\dots,2L^2$ for honeycomb lattice.

While pinning is typically positioned at the system edges, i.e., at $i=1$, $i=N$,
the last $L-1$ ($2L-1$) sites at both ends of projected 1D chain corresponding
to square (honeycomb) lattice are left without their original bonds, necessitating
the placement of pinnings at these outer sites to accurately estimate staggered
magnetization [see Fig.\ref{fig:lattice}(c)]. Therefore, we define the magnitude
of staggered magnetization for a given spin $S$ and XXZ anisotropy $\Delta$ as
\begin{align}
m_\alpha^z(S,\Delta)=\lim_{N \to \infty}|\braket{S^{z}_{N/2}}-\braket{S^{z}_{N/2+1}}|/2.
\label{eq:mst}
\end{align}
Examples of local spin moment profiles, $\braket{S^z_i}$, for the $S^z=0$ sector
at $\Delta=1$ are depicted in the top panels of Fig.\ref{fig:localSz}.
Here, the parameter $\alpha$ adopts the value `sq' for
square lattices and `hon' for honeycomb lattices. In both square
and honeycomb lattices, the oscillation of $\braket{S^z_i}$ smoothly decays towards
the center of the system, validating the approach of defining the N\'eel order
parameter via the local moments of the two central spins, $\braket{S^{z}_{N/2}}$
and $\braket{S^{z}_{N/2+1}}$.

\subsubsection{spin gap}

The spin gap offers insight into phenomena such as quantum
fluctuations and the stability of the N\'eel state. It also provides
us with understanding of how classical behavior emerges from quantum
systems. The spin gap (singlet-triplet gap)  for given $S$ and $\Delta$
is defined as follows:
\begin{align}
\varepsilon_\alpha(S,\Delta)=\lim_{N \to \infty} [E_0(N,1)-E_0(N,0)],
\label{eq:gap}
\end{align}
where $E_0(N,S^z)$ is the total ground-state energy
of the system with $N$ sites and the $z$-component of the total
spin, $S^z$. Here, the parameter $\alpha$ adopts the value `sq' for
square lattices and `hon' for honeycomb lattices.  

To verify the validity of our spin gap calculations under the imposed pinning
distribution, we examine the spatial distribution of a spinon as the spin sector
transitions from $S^z=0$ to $S^z=1$. The spatial distribution of the spinon is
visualized by seeing the variance in $\braket{S^z_i}$ between the $S^z=0$ and
$S^z=1$ sectors, $\delta \braket{S^z_i}$. We plot the profile of the spinon
distribution for $\Delta=1$ in the bottom panels of Fig.\ref{fig:localSz}.
In both square and honeycomb lattices, it is observed that the probability of
spinon presence is maximized near the center of the system, with minimal
presence near the edges. This observation confirms that the spin gap in the
bulk limit can be correctly estimated using Eq.~(\ref{eq:gap}). We note that 
the distribution of $\braket{S^z_i}$ for the $S^z=1$ sector and $\delta \braket{S^z_i}$
are aymmetric because the rotational symmetry as well as mirror symmetry
of the system is broken.

\section{Results}

\subsection{staggered magnetization}

We start by looking at what we found about staggered magnetization. This measure adeptly quantifies the AFM alignment of magnetic moments throughout the lattice, offering profound insights into the stability and resilience of the N\'eel state against external perturbations. Our investigation spans two lattice configurations, i.e., square and honeycomb structures, and encompasses systems with spin magnitudes of $S=1/2$ and $S=1$. The primary objective of this analysis is to delineate the degree to which staggered magnetization is influenced by the lattice geometry and spin values.

\subsubsection{square lattice}

\begin{figure}[t]
	\centering
	\includegraphics[width=1.0\linewidth]{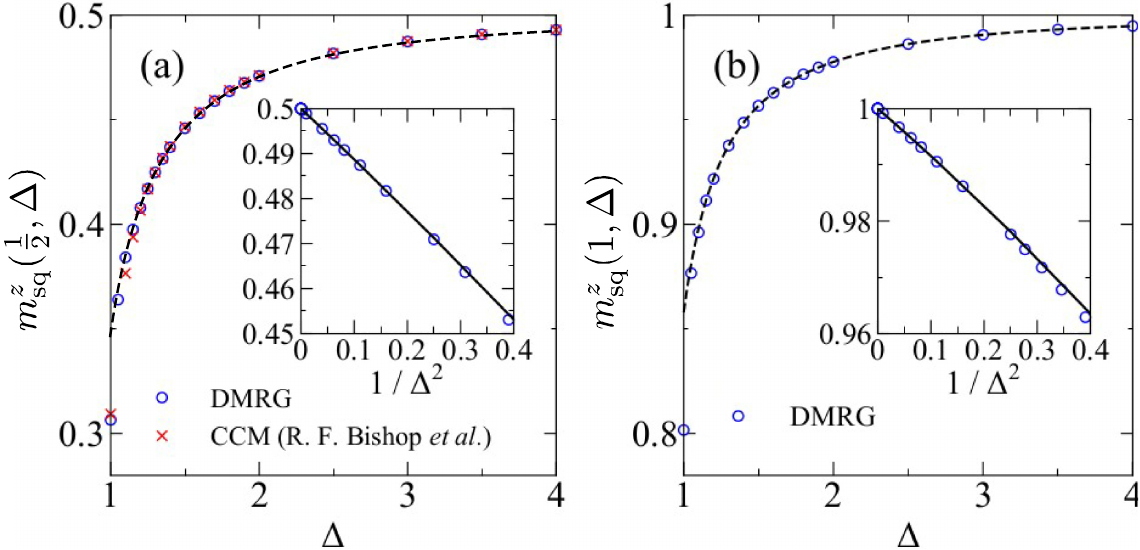}
	\caption{
		Extrapolated values of $m_{\mathrm{sq}}^z(S,\Delta)$
		to the thermodynamic limit as a function of $\Delta$ for the (a) $S=1/2$ and
		(b) $S=1$ square-lattice XXZ models. Dashed lines represent the results
		from the SE up to the 12th order $1/\Delta$~\cite{Singh1989,Singh1990,Weihong1991-1}. Red crosses denote
		the results obtained via the CCM~\cite{BISHOP2017} for the $S=1/2$ case.
		Insets show $m_{\mathrm{sq}}^z(S,\Delta)$ versus $1/\Delta^2$ in the
		large-$\Delta$ region, with solid lines indicating the SE results up to the 6th
		order of $1/\Delta$.
	}
	\label{fig:mz_square}
\end{figure}

In our preceding study~\cite{Kadosawa2023-1}, the magnitude of
staggered magnetization for the $S=1/2$ and $S=1$ square-lattice XXZ
Heisenberg models was quantified as a function of $\Delta$. 
We here revisit these results to consider the extent of quantum fluctuations
in the N\'eel state at $\Delta \ge 1$. They are plotted in
Fig.~\ref{fig:mz_square}. For any $S$, the quantum fluctuations
are maximized at $\Delta=1$, correspondingly minimizing the staggered
magnetization magnitude. Specifically, for $S=1/2$, we previously estimated
$m^z_{\rm sq}(\frac{1}{2},1.0)=0.3065$~\cite{Kadosawa2023-1}, aligning
closely with extant numerical estimates: $0.3067$ by DMRG~\cite{White2007},
$0.3093$ via the coupled cluster method (CCM)~\cite{BISHOP2017}, and
$0.30743$ through quantum Monte Carlo (QMC)~\cite{Sandvik2010}. 
These values approximate $60\%$ of the classical one
$m^z_{\rm sq}(\frac{1}{2},\infty)=0.5$. For $S=1$, the staggered magnetization
rises to about $80\%$ of its classical counterpart $m_z^{\rm st}(1,\infty)=1$,
as numerically determined to be $m^z_{\rm sq}(1,1.0)=0.8017$~\cite{Kadosawa2023-1}
via DMRG and $m^z_{\rm sq}(1,1.0)=0.802$~\cite{Niesen2017} by infinite
projected entangled pair states (iPEPS). These results underscore
the substantial suppression of quantum fluctuations with an increase in $S$.
In fact, an expansion in terms of
$1/S$ yields $m^z_{\rm sq}(S,1.0) = 1 - 0.1966019S^{-1} + 0.00087S^{-3} + O(S^{-4})$~\cite{Hamer1992,Igarashi1992,Canali1993}.
Since the coefficients of higher order terms than $1/S$ are
very small, a rapid convergence to the classical (or Ising) limit 
$m^z_{\rm sq}(S,1.0)/S\to1$ with increasing $S$ is naively
expected. This trend has received numerical validation~\cite{Kadosawa2023-1}.

To explore the evolution of quantum fluctuations with $\Delta$,
we compare our DMRG results to SE predictions. The SE results are
plotted by dotted lines in Fig.~\ref{fig:mz_square}. At $\Delta=1$,
our DMRG analysis yields $m^z_{\rm sq}(\frac{1}{2},1.0)=0.3065$ and
$m^z_{\rm sq}(1,1.0)=0.8017$, whereas SE up to the 12th order offers
$m^z_{\rm sq}(\frac{1}{2},1.0)=0.3462$ and $m^z_{\rm sq}(1,1.0)=0.8579$.
Despite incorporating terms up to the 12th order, a discrepancy remains
due to unaccounted quantum fluctuations, with errors around $7.9\%$
for $S=1/2$ and $5.6\%$ for $S=1$. Increasing $\Delta$ slightly to
$1.05$ reduces these deviations significantly to $0.76\%$ and $0.44\%$
for $S=1/2$ and $S=1$, respectively, illustrating a marked decrease in quantum fluctuations with enhanced XXZ anisotropy. In the $\Delta=\infty$ limit, SE approaches exactness, nullifying quantum fluctuations. These discrepancies highlight the challenge of fully
accounting for quantum fluctuations, with a deviation of approximately $7.9\%$ for
$S=1/2$ and $5.6\%$ for $S=1$. However, a slight increase in $\Delta$ to $1.05$ yields 
$m^z_{\rm sq}(\frac{1}{2},1.05)=0.3640$ (DMRG) and
$m^z_{\rm sq}(\frac{1}{2},1.05)=0.3678$ (SE12) for $S=1/2$, and 
$m^z_{\rm sq}(1,1.05)=0.8767$ (DMRG) and
$m^z_{\rm sq}(1,1.05)=0.8811$ (SE12) for $S=1$,
significantly reducing deviations to $0.76\%$ and $0.44\%$, respectively.
(Hereinafter, when considering up to the $n$th order in the SE,
we denote it as SE$n$ if needed.)
This definitely suggests a pronounced reduction in quantum fluctuations
attributable to XXZ anisotropy.

In the limit of $\Delta=\infty$, the SE analysis approaches exact
since quantum fluctuations disappear. Utilizing SE up to the 6th order in
$1/\Delta$, we express the staggered magnetization for $S=1/2$
systems as
$2m^z_{\rm sq}(\frac{1}{2},\Delta) = 1 + m_2/\Delta^2 + m_4/\Delta^4 + m_6/\Delta^6$.
The coefficients are calculated to be $m_2 = -2/9 = -0.222222\dots$,
$m_4 = -8/255 = -0.0355555\dots$, and $m_6 = -0.01894258$ for
$S=1/2$~\cite{Davis1960,Huse1988,Parrinello1974,Singh1989,Weihong1991-1}.
In the case of $S=1$ systems, staggered magnetization is similarly formulated as
$m^z_{\rm sq}(1,\Delta) = 1 + m_2/\Delta^2 + m_4/\Delta^4 + m_6/\Delta^6$,
with coefficients $m_2 = -4/49 = -0.081632653\dots$, $m_4 = -0.026959099$, and
$m_6 = -0.0136997515$~\cite{Singh1990,Weihong1991-1}.
The magnitude of the lowest-order term, i.e., $m_2$, for $S=1/2$ is approximately 2.7
times that for $S=1$, implying a significant difference in quantum fluctuations between
the two. Fitting our data for $0 \le 1/\Delta \le 0.05$, we obtain $m_2= -0.222222225$,
$m_4= -0.0355542736$, and $m_6=-0.0189663810$ for $S=1/2$, and
$m_2=-0.081632653$, $m_4=-0.026959397$, and $m_6=-0.013867377$ for $S=1$.
These coefficients are in almost perfect agreement with the SE results.

Additionally, in contrast to the singular behavior near $\Delta=1$ predicted by
SWT, which posits that $m^z_{\rm sq}(\frac{1}{2},\Delta)=\sum_{n=0}^{\infty}m_n(\Delta-1)^{n/2}$~\cite{Weihong1991-1,Weihong1991-2,Huse1988}, our observations reveal that $m^z_{\rm sq}(\frac{1}{2},\Delta)$
is almost linearly proportional to $\Delta-1$ within the range $1\le\Delta\lesssim1.01$.
This behavior aligns with results from the CCM)~\cite{BISHOP2017}. However,
we further ascertain that, except in the immediate vicinity of $\Delta=1$, the staggered
magnetization for the square-lattice XXZ Heisenberg model at $\Delta \gtrsim 1$ can
be approximately and quantitatively captured by SE, provided the expansion includes
up to the 12th order of $1/\Delta$.

\subsubsection{honeycomb lattice}

\begin{figure}[t]
	\centering
	\includegraphics[width=1.0\linewidth]{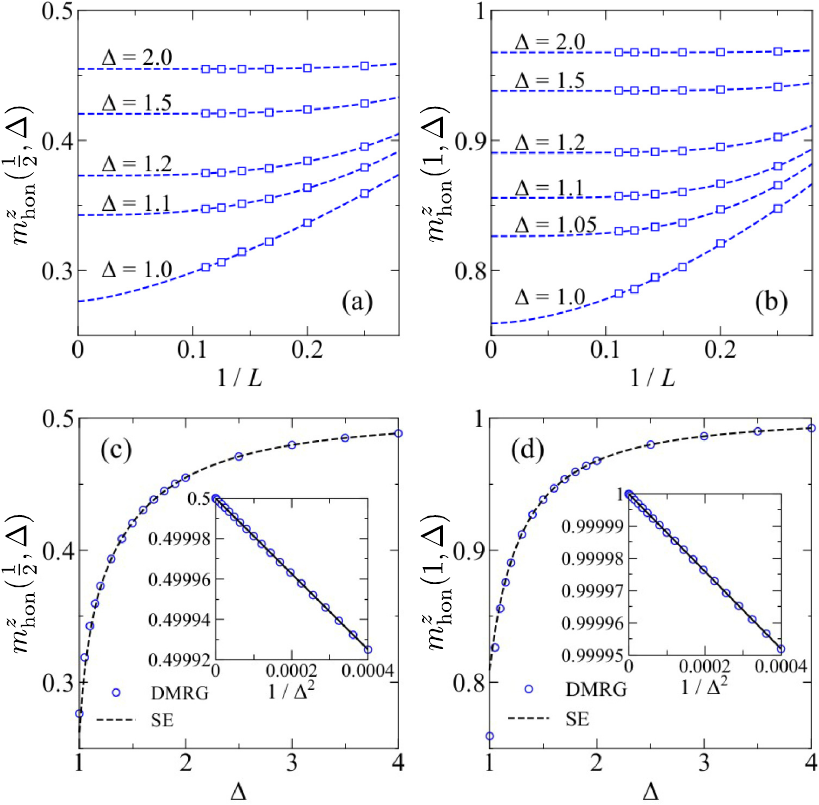}
	\caption{
		Magnitude of staggered magnetization for the $S=1/2$ and $S=1$
		honeycomb-lattice XXZ models as a function of $\Delta$. Finite-size scaling
		analyses of (a) $m^z_{\rm hon}(\frac{1}{2},\Delta)$ and
		(b) $m^z_{\rm hon}(1,\Delta)$. Extrapolated values of (c) $m^z_{\rm hon}(\frac{1}{2},\Delta)$ and (d) $m^z_{\rm hon}(1,\Delta)$ to
		the thermodynamic limit as a function of $\Delta$. The dashed lines
		show the SE results up to the 12th order of $1/\Delta$~\cite{Oitmaa1992}.
		Insets show $m^z_{\rm hon}(S,\Delta)$ versus $1/\Delta^2$
		in the large-$\Delta$ region. Solid lines represent the SE results
		up to the 6th order of $1/\Delta$.
	}
	\label{fig:mz_honeycomb}
\end{figure}

In exploring the honeycomb-lattice XXZ model, we note that each site is
connected to three neighboring sites, a contrast to the four neighbors in
the square-lattice model. This difference leads us to hypothesize that the
N\'eel state in the honeycomb-lattice model might exhibit less stability
compared to its counterpart in the square model due to the potential for
increased quantum fluctuations. To investigate this, we delve into the
dependence of staggered magnetization on the anisotropy parameter $\Delta$
for both $S=1/2$ and $S=1$ within the honeycomb lattice, drawing comparisons
with our previous findings for the square-lattice model.

In Figs.~\ref{fig:mz_honeycomb}(a,b), we perform finite-size scaling analyses
for $m^z_{\mathrm{hon}}(\frac{1}{2},\Delta)$ and $m^z_{\mathrm{hon}}(1,\Delta)$
over various values of $\Delta$. Broadly, this scaling examines the decay of Friedel
oscillations with distance from a spin pinned at the system edge. Given that
the N\'eel state features a staggered (commensurate) arrangement of spins in
our projected 1D chain, we anticipate a smooth decay of the Friedel oscillations
with distance. This has been verified in Sec.~\ref{subsec:stagmag}.
Consequently, $m^z_{\mathrm{hon}}(\frac{1}{2},\Delta)$ and
$m^z_{\mathrm{hon}}(1,\Delta)$ extrapolate smoothly to the thermodynamic limit
as functions of inverse system size. Furthermore, the observation that convergence
with respect to size occurs more rapidly as $\Delta$ increases reflects the diminishing
quantum fluctuations.

The extrapolated values of $m^z_{\mathrm{hon}}(\frac{1}{2},\Delta)$ and
$m^z_{\mathrm{hon}}(1,\Delta)$ are plotted as a function of $\Delta$ in
Figs.~\ref{fig:mz_honeycomb}(c,d), respectively. In the case where
the quantum fluctuations are largest at $\Delta=1$, we obtain
$m^z_{\mathrm{hon}}(\frac{1}{2},1.0)=0.2764$ for $S=1/2$ and
$m^z_{\mathrm{hon}}(1,1.0)=0.7646$ for $S=1$. These values are in close
agreement with various numerical methods from previous studies:
$0.2857$~\cite{Ganesh2013},
$0.2720$~\cite{Zhu2013},
$0.2611$~\cite{Gong2013} by DMRG,
$0.2677$~\cite{Castro2006} by QMC,
$0.262$~\cite{Albuquerque2011} by exact diagonalization,
$0.2730$~\cite{Bishop2016} by CCM for $S=1/2$,
and $0.7412$~\cite{Bishop2016} by CCM for $S=1$.
These are $\sim55\%$ and $\sim76\%$ of their respective classical values, only
slightly smaller despite of the greater quantum fluctuations compared to
the square lattice case. On the other hand, the SE analyses up to the 12th order
of $1/\Delta$ lead to $m^z_{\rm hon}(\frac{1}{2},1)=0.3409$ and
$m^z_{\rm hon}(1,1)=0.8139$ for $S=1/2$ and $S=1$, respectively. These values,
when compared with those obtained from our DMRG simulations, exhibit
discrepancies of $6.5\%$ for $S=1/2$ and $4.9\%$
for $S=1$. Interestingly, these deviations are rather smaller than those observed for
the square lattice, where the discrepancies are notably lower at $7.9\%$ for $S=1/2$
and  $5.6\%$ for $S=1$. Furthermore, the increase in magnetization when
$S$ changes from $1/2$ to $1$ is similar to that in the square lattice.
This may be expected from the coefficients of the $1/S$ series expansion for
the honeycomb lattice
$m^z_{\mathrm{hon}}(S,1.0)=1-0.258193/S+\cdots$~\cite{Weihong1991-2},
which are close to those for the square lattice.

Let us then see the $\Delta$-dependence. As illustrated in Figs.~\ref{fig:mz_honeycomb}(c,d), the magnetization rapidly approaches classical values with increasing $\Delta$. It is also evident that, apart from the immediate vicinity of $\Delta=1$, the magnetization is well captured by the SE12 predictions. Indeed, a slight increase in $\Delta$ from $1$ to $1.05$ yields $m^z_{\text{hon}}(\frac{1}{2},1.05)=0.3188$ (DMRG) versus $m^z_{\text{hon}}(\frac{1}{2},1.05)=0.3478$ (SE12) for $S=1/2$, and $m^z_{\text{hon}}(1,1.05)=0.8242$ (DMRG) versus $m^z_{\text{hon}}(1,1.05)=0.8427$ (SE12) for $S=1$. The discrepancies between the DMRG and SE12 values significantly decrease from $6.5\%$ and $4.9\%$ at $\Delta=1$ to $2.9\%$ and $1.9\%$ at $\Delta=1.05$ for $S=1/2$ and $S=1$, respectively. Nevertheless, the quantum fluctuations remain comparatively large, and thus the reduction is not as dramatic as in the case of the square lattice, where the discrepancies decreased from $7.9\%$ and $5.6\%$ to $0.76\%$ and $0.44\%$.

By employing expansions up to the 4th order in $1/\Delta$, the staggered magnetization for both spin-$1/2$ and $S=1$ systems can be expressed as $m_z(S=1/2)=0.5-m_2/\Delta^2+m_4/\Delta^4+o(1/\Delta^6)$. The coefficients are found to be $m_2 = -3/16 = -0.1975$, $m_4 = 31/768 =0.0403645833\dots$ for $S=1/2$ and $m_2 = -3/25=-0.12$, $m_4 = -17977/54000=-0.3329074\dots$ for $S=1$~\cite{Oitmaa1992}. Fitting our data for $0 \le 1/\Delta \le 0.02$, we obtain
$m_2=-0.187499964$, $m_4= 0.0394523682$ for $S=1/2$ and $m_2=-0.120000004$,
$m_4=-0.0332430099$ for $S=1$. These values are in good agreement with those
from SE.

Thus, our examination of the honeycomb-lattice XXZ model reveals that, akin to the case for square lattice, staggered magnetization can be reasonably approximated by SE12 for $\Delta > 1$.




\subsection{Spin gap}

Next, we delve into the investigation of the spin gap, which serves as an indicator
of the stability of N\'eel LRO when spin rotation symmetry about the z-axis is explicitly
broken by staggered magnetization. This parameter is essential for understanding
the energy required to excite the system from its N\'eel ground state, a facet less
explored compared to direct magnetization measurements. Employing methodologies
analogous to those used in our magnetization studies, we extend our analysis to
both square and honeycomb lattice configurations for $S=1/2$ and $S=1$ systems.
Our aim is to elucidate the behavior of the spin gap across varying lattice geometries
and spin magnitudes, offering insights into the intricacies associated with excitations
from the N\'eel state.

\subsubsection{square lattice}

\begin{figure}[t]
	\centering
	\includegraphics[width=1.0\linewidth]{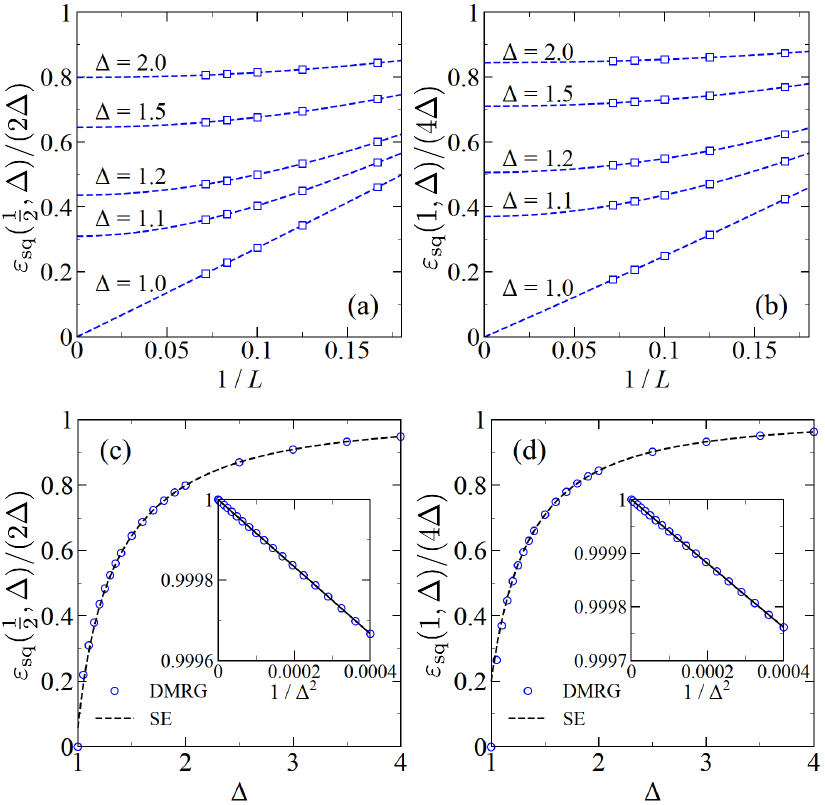}
	\caption{
		Spin gap for the $S=1/2$ and $S=1$ square-lattice
		XXZ models as a function of $\Delta$.
		Finite-size scaling analyses of
		(a) $\varepsilon_{\mathrm{sq}}(\frac{1}{2},\Delta)$ and
		(b) $\varepsilon_{\mathrm{sq}}(1,\Delta)$.
		Extrapolated values of
		(c) $\varepsilon_{\mathrm{sq}}(\frac{1}{2},\Delta)$ and
		(d) $\varepsilon_{\mathrm{sq}}(1,\Delta)$.
		to the thermodynamic limit as a function of $\Delta$.
		The dashed lines show the SE results up to the 10th order of
		$1/\Delta$~\cite{Weihong1991-1}.
		Insets show $\varepsilon_{\mathrm{sq}}(S,\Delta)$ in the
		large $\Delta$ region.Solid lines represent the SE results
		up to the 6th order of $1/\Delta$.
	}
	\label{fig:gapsq}
\end{figure}

Let us first examine the spin gap in the case of a square lattice.
In Figs.~\ref{fig:gapsq}(a,b), finite-size scaling analyses for
$\varepsilon_{\mathrm{sq}}(\frac{1}{2},\Delta)$ and
$\varepsilon_{\mathrm{sq}}(1,\Delta)$ are shown across various values
of $\Delta$. Our scaling analysis reveals that a smooth extrapolation
of the spin gap often suggests that the scaling function resembles
the contour of magnon band near the Fermi level. For both spin
magnitudes at $\Delta=1$, the spin gap data aligns closely with
a linear fit, extrapolating towards zero in the thermodynamic limit,
albeit with minor numerical uncertainties inherent to the extrapolation
process. This linear fit aligns with expectations for gapless systems
where the linear magnon band structure near the Fermi points
dominates. The actual values of extrapolated spin gap are $\varepsilon_{\mathrm{sq}}(\frac{1}{2},1)=-0.00170006$ and $\varepsilon_{\mathrm{sq}}(1,1)=0.00684318$. This hints at
a Néel LRO that, despite being stable, exhibits a spin gap of
zero due to the arbitrary direction of symmetry breaking.
Moreover, at slightly increased values of $\Delta$, namely
$1.1$ and $1.2$, the observed small gaps corroborate the
quadratic dispersion expected near the Fermi points.
As the gap widens, quantum fluctuations wane, leading to
a narrower bandwidth and, thus, a diminished size dependence
of the gap. Utilizing SBC to transform the 2D lattice into
an effective 1D system allows for the original 2D Fermi
surface to be conceptualized as a 'Fermi line,' aiding in
the smooth scaling of the gap.

Figs.~\ref{fig:gapsq}(c,d) display the extrapolated spin gaps $\varepsilon_{\mathrm{sq}}(\frac{1}{2},\Delta)$ and $\varepsilon_{\mathrm{sq}}(1,\Delta)$, showcasing a trend similar
to that of magnetization with increasing $\Delta$, rapidly approaching
the classical values $4\Delta S$. Excluding the region immediately
around $\Delta=1$, the behavior of the spin gap correlates well
with SE predictions up to the 10th order in $1/\Delta$. 
Specifically, a SE up to the 2nd order in $1/\Delta$ formulates
$\varepsilon_{\mathrm{sq}}(S,\Delta)=2+m_2/\Delta^2+{\cal O}(1/\Delta^4)$,
where $m_2$ equals $-5/3=1.6666666\cdots$ for $S=1/2$ and
$-50/21=2.3809523\cdots$ for $S=1$. Our data fitting for $0 \le 1/\Delta \le 0.02$
yields $m_2=-1.66803135$ for $S=1/2$ and $m_2=-2.37996804$ for $S=1$,
in close agreement with these SE coefficients, demonstrating the validity of
our method.

\begin{figure}[t]
	\centering
	\includegraphics[width=1.0\linewidth]{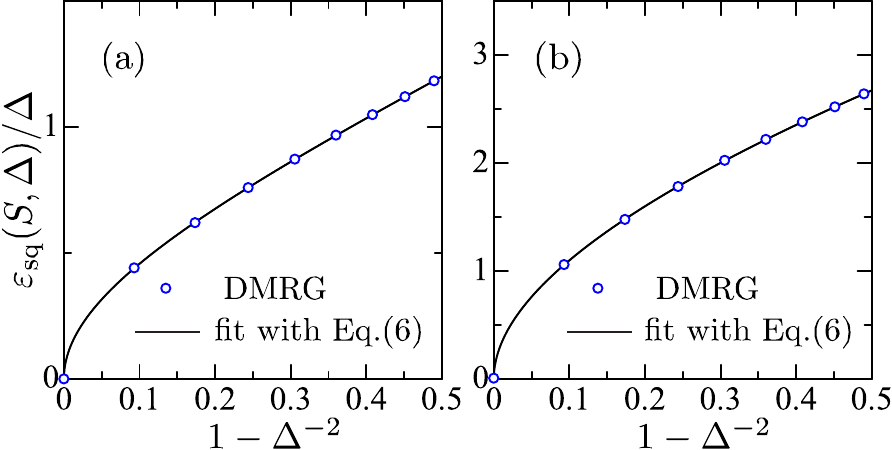}
	\caption{
		$\Delta$-normalized spin gap for the (a) $S=1/2$ and (b) $S=1$ square-lattice
		XXZ models as a function of $1-\Delta^{-2}$. Circles denote DMRG results
		in the thermodynamic limit, while the solid line represents a fit using
		Eq.~(\ref{eq:singularity}).
	}
	\label{fig:gapsqexponent}
\end{figure}

\begin{table}
	\caption{
		Comparison of spin gaps as a function of $\Delta$ for the $S=1/2$ square-lattice XXZ Heisenberg model obtained via DMRG and CCM calculations.
		}
	\begin{center}
		\begin{tabular}{wc{1.0cm}wc{1.4cm}wc{1.4cm}|wc{1.0cm}wc{1.4cm}wc{1.4cm}}
			\hline
			\hline
			\multirow{2}{*}{$\Delta$} &  \multicolumn{2}{c}{$\varepsilon_{\mathrm{sq}}(\frac{1}{2},\Delta)$} & \multirow{2}{*}{$\Delta$} &  \multicolumn{2}{c}{$\varepsilon_{\mathrm{sq}}(\frac{1}{2},\Delta)$ } \\
			&  DMRG  & CCM &   & DMRG  &  CCM \\
			\hline
			1.00 &  -0.0017  &  -0.0086   &  1.60   &  2.1995  &  2.2279  \\
			1.10 &   0.6810  &   0.5601   &  1.70   &  2.4578  &  2.4921  \\
			1.15 &   0.8722  &   0.7811   &  1.80   &  2.7083  &  2.7465  \\
			1.20 &   1.0461  &   0.9805   &  1.90   &  2.9535  &  2.9934  \\
			1.25 &   1.2088  &   1.1646   &  2.00   &  3.1942  &  3.2344  \\
			1.30 &   1.3627  &   1.3371   &  2.50   &  4.3496  &  4.3828  \\
			1.35 &   1.5118  &   1.5004   &  3.00   &  5.4546  &  5.4790  \\
			1.40 &   1.6560  &   1.6563   &  3.50   &  6.5306  &  6.5481  \\
			1.50 &   1.9334  &   1.9509   &  4.00   &  7.5880  &  7.6008  \\
			\hline                                                
			\hline
		\end{tabular}
	\end{center}
	\label{table:sqgapS12}
\end{table}

Investigating the behavior of spin gap near $\Delta=1$, SWT anticipates
a singularity, encapsulated by the following relation:
\begin{align}
	\frac{\varepsilon}{\Delta}=\eta_1(1-\Delta^{-2})^{1/2}+\eta_2(1-\Delta^{-2})+\eta_1(1-\Delta^{-2})^{3/2}+\cdots,
	\label{eq:singularity}
\end{align}
with constants $\eta_1=4S-0.431436$ and $\eta_2=1.2732$ derived
from SWT predictions~\cite{Hamer1991}. To scrutinize this predicted
behavior, we present DMRG results for the normalized spin gap,
$\frac{\varepsilon}{\Delta}$, as a function of $1-\Delta^{-2}$ around
$\Delta=1$ in Figs.~\ref{fig:gapsqexponent}(a,b). For both $S=1/2$ and
$S=1$, we can reasonably fit our DMRG data by Eq.~(\ref{eq:singularity}),
yielding $\eta_1=1.36394987$, $\eta_2=0.0667245772$, $\eta_3=0.572388620$
for $S=1/2$ and $\eta_1=3.23450360$, $\eta_2=0.754467400$,
$\eta_3=0.0226881660$ for $S=1$. These leading coefficients are
close to the SWT prediction of $\eta_1=1.568564$ for $S=1/2$ and
$\eta_1=3.568564$ for $S=1$, indicating an increase in $\eta_1$ with $S$,
consistent with the fact that in the large $S$ limit, the gap jumps to
$\varepsilon_{\mathrm{sq}}(S=\infty,\Delta=1^+)=4 \Delta S$ as soon as XXZ anisotropy is introduced. Our numerical investigation thus substantiates
the SWT-predicted singularity in the spin gap near $\Delta=1$. Nonetheless,
for $S=1/2$, contrasting viewpoints emerge, such as those from CCM
analyses, which suggest a near-linear relation, $\varepsilon \propto \Delta$,
diverging from the expected singular behavior. This discrepancy could arise
from our extrapolations to the thermodynamic limit, particularly near
$\Delta=1$, which are marginally higher than those deduced from CCM.
As reference, Table~\ref{table:sqgapS12} compares the spin gap values
derived from DMRG and CCM for various $\Delta$ settings in the $S=1/2$
case. A more in-depth examination will be imperative in future studies to
resolve these discrepancies and fully delineate the characteristics of spin gap
near $\Delta=1$.

\subsubsection{honeycomb lattice}

\begin{figure}[t]
	\centering
	\includegraphics[width=1.0\linewidth]{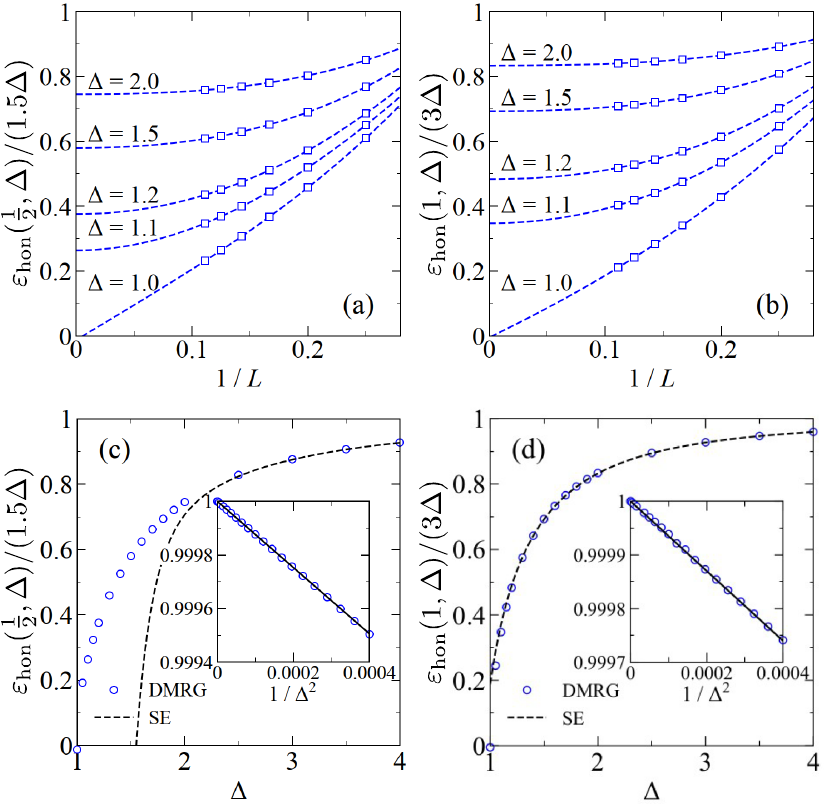}
	\caption{
		Spin gap for the $S=1/2$ and $S=1$ honeycomb-lattice
		XXZ models as a function of $\Delta$. Finite-size scaling analyses of
		(a) $\varepsilon_{\mathrm{hon}}(\frac{1}{2},\Delta)$ and
		(b) $\varepsilon_{\mathrm{hon}}(1,\Delta)$.
		Extrapolated values of
		(c) $\varepsilon_{\mathrm{hon}}(\frac{1}{2},\Delta)$ and
		(d) $\varepsilon_{\mathrm{hon}}(1,\Delta)$.
		to the thermodynamic limit as a function of $\Delta$.
		The dashed lines show the SE results up to the 10th order of
		$1/\Delta$~\cite{Oitmaa1992}.
		Insets show $\varepsilon_{\mathrm{sq}}(S,\Delta)$ in the
		large $\Delta$ region.Solid lines represent the SE results
		up to the 6th order of $1/\Delta$.
	}
	\label{fig:gaphon}
\end{figure}

Finally, we examine the spin gap for the honeycomb-lattice model.
In Figs.~\ref{fig:gaphon}(a,b), we conduct finite-size scaling analyses for
$\varepsilon_{\mathrm{hon}}(\frac{1}{2},\Delta)$ and $\varepsilon_{\mathrm{hon}}(1,\Delta)$
over various values of $\Delta$. We see that smooth scaling is possible
for all values of $\Delta$, similar to the case of the square lattice.
Consistently, at $\Delta=1$, the scaling function is almost linear,
approaching to nearly zero as $1/L$ decreases; while for $\Delta>1$,
a quadratic behavior indicative of gap opening is observed. The actual
extrapolated values are $\varepsilon_{\mathrm{hon}}(\frac{1}{2},1)=-0.01916648$
and $\varepsilon_{\mathrm{hon}}(1,1)=-0.01447616$, indicating slight
but larger deviations from zero than those observed in the square-lattice
case. The square lattice, having one site per structural unit cell, allowed
calculations up to $L=14$, whereas the honeycomb lattice, with two sites
per structural unit cell, limits computations to $L=9$ for a comparable
computational cost, potentially leading to relatively larger scaling errors
towards the thermodynamic limit in the honeycomb-lattice case.

The extrapolated values of $\varepsilon_{\mathrm{hon}}(\frac{1}{2},\Delta)$
and $\varepsilon_{\mathrm{sq}}(1,\Delta)$ are plotted as a function of $\Delta$
in Figs.~\ref{fig:gaphon}(c,d), showing a behavior broadly similar to
the magnetization versus $\Delta$ relationship. Interestingly, the saturation
towards the classical value $3\Delta S$ seems to be more gradual than
observed in the square-lattice case, possibly a reflection of heightened
quantum fluctuations within the honeycomb structure. 

A notable finding is that SE analyses up to the 10th order in $1/\Delta$
are almost inapplicable in the region $\Delta \lesssim 2$ for S=1/2.
Due to the large coefficients of higher-order terms, adding each successive
term causes the values to oscillate significantly near $\Delta=1$, making
approximation by SE very hard in the vicinity of $\Delta=1$ (see Appendix A).
This emphasizes the critical influence of quantum fluctuations on the spin
excitation in its N\'eel state. Conversely, for $S=1$, the SE results up to the
10th order in $1/\Delta$ can reasonably describe the general gap behavior,
except in the immediate vicinity of $\Delta=1$.

To substantiate the accuracy of our DMRG calculations, we undertake
a comparison with SE predictions, particularly in the regime of large $\Delta$.
The SE up to the 2nd order in $1/\Delta$ is expressed as
$\varepsilon_{\mathrm{hon}}(S,\Delta) = 2 + m_2/\Delta^2 + {\cal O}(1/\Delta^4)$,
where the coefficients are determined to be $m_2 = -15/8 = 1.875$ for $S=1/2$
and $m_2 = -39/20 = 1.95$ for $S=1$~\cite{Oitmaa1992}. Fitting our DMRG data
within the range $0 \le 1/\Delta \le 0.02$ yields coefficients $m_2 = -1.85284825$
for $S=1/2$ and $m_2 = 1.94992536$ for $S=1$. These findings are in remarkable
concordance with the established SE coefficients, further affirming the reliability
of our DMRG computational approach.

\begin{figure}[t]
	\centering
	\includegraphics[width=1.0\linewidth]{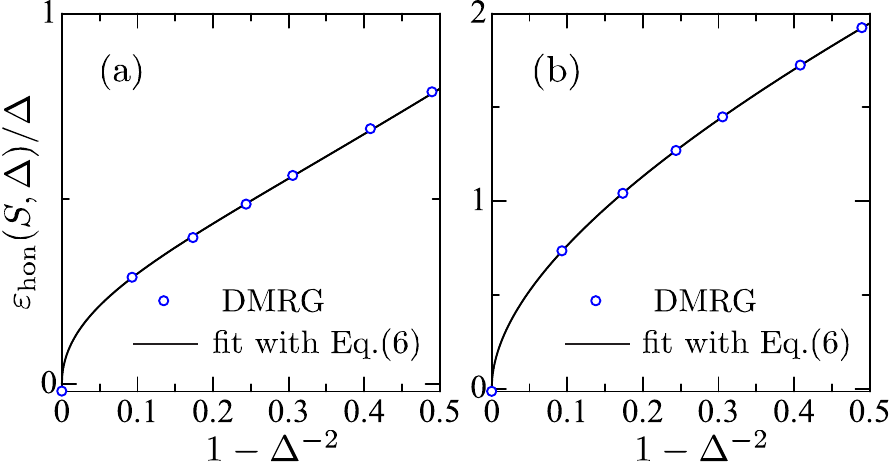}
	\caption{
		$\Delta$-normalized spin gap for the (a) $S=1/2$ and (b) $S=1$
		honeycomb-lattice XXZ models as a function of $1-\Delta^{-2}$.
		Circles denote DMRG results in the thermodynamic limit, while the solid line
		represents a fit using Eq.~(\ref{eq:singularity}).
	}
	\label{fig:gaphonexponent}
\end{figure}

In alignment with observations for the square lattice, SWT also forecasts
singular behavior near $\Delta=1$ for the honeycomb lattice. 
The asymptotic form of this behavior is encapsulated by
Eq.~(\ref{eq:singularity}), with SWT providing the coefficients as
$\eta_1=3S-0.423239$ and $\eta_2=1.2405$\cite{Weihong1991-2}.
To verify if our DMRG data align with these predictions, we analyze
the normalized spin gap, $\frac{\varepsilon}{\Delta}$, as a function of
$1-\Delta^{-2}$ approaching $\Delta=1$, as depicted in
Figs.~\ref{fig:gaphonexponent}(a,b).

Our fits to Eq.~(\ref{eq:singularity}) for both $S=1/2$ and $S=1$
seem to be reasonable and yield $\eta_1=1.03093109$, $\eta_2=-0.614589918$,
$\eta_3=1.06200153$ for $S=1/2$, and $\eta_1=2.12914213$,
$\eta_2=0.923738066$, $\eta_3=-0.0550192661$ for $S=1$.
These leading coefficients exhibit a notable correspondence with
SWT-anticipated $\eta_1=1.076761$ for $S=1/2$ and $\eta_1=2.576761$
for $S=1$, underscoring our numerical validation of the predicted singularity
in the spin gap as $\Delta$ approaches $1$. Incidentally, the observation
that $\eta_1$ for $S=1$ is larger than that for $S=1/2$ suggests a trend
towards the discontinuity where the gap leaps to
$\varepsilon_{\mathrm{sq}}(S=\infty,\Delta=1^+)=3 \Delta S$ with
the introduction of XXZ anisotropy in the limit of $S=\infty$.

\section{summary and discussion}

\begin{figure}[t]
	\centering
	\includegraphics[width=0.7\linewidth]{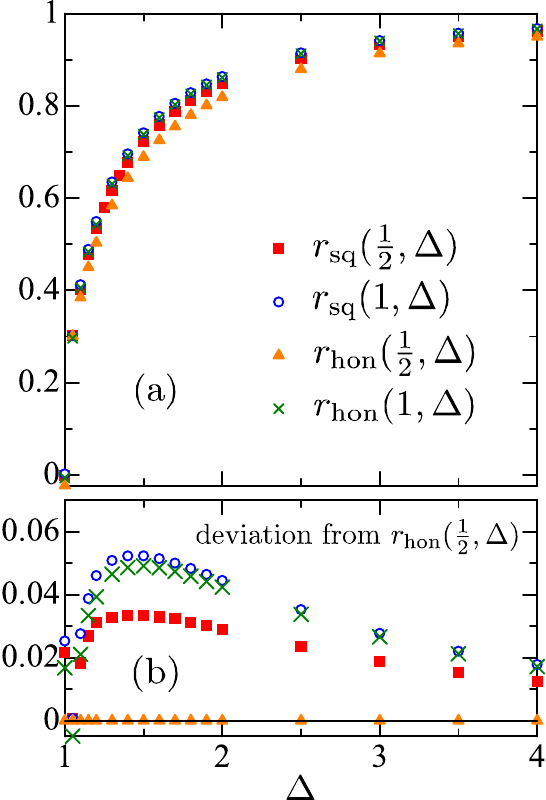}
	\caption{
		(a) Ratio of the spin gap to the magnitude of magnetization, which is normalized to approach unity in the classical limit, as a function of $\Delta$ for the the
		$S=1/2$ and $S=1$ XXZ Heisenberg models on square and honeycomb lattices.
		(b) The values plotted in (a) subtracted by $r_{\mathrm{hon}}(\frac{1}{2},\Delta)$.
	}
	\label{fig:ratio}
\end{figure}

We have achieved a comprehensive study of the $S=1/2$ and $S=1$ XXZ
Heisenberg model on square and honeycomb lattices. By employing the
DMRG method, we systematically analyzed the evolution of staggered
magnetization and the associated spin gap across a wide range of easy-axis
anisotropies. A key to enhancing DMRG performance was the implementation
of SBC, which enabled an exact mapping of the original 2D clusters onto
a 1D chain. This technique significantly improved our ability to perform
efficient finite-size scaling analysis, thereby facilitating the extrapolation
of physical quantities to the thermodynamic limit with high accuracy. 

Given the difference in the number of adjacent sites -- four for the square lattice
versus three for the honeycomb lattice -- it is reasonable to anticipate greater
quantum fluctuations in the N\'eel phase for the honeycomb structure. This implies
a potentially lower stability of the N\'eel LRO in the honeycomb lattice as compared
to the square lattice. Contrary to what might be expected from the increased
quantum fluctuations, we found that the magnitude of staggered magnetization
in the honeycomb lattice is only marginally smaller than that in the square lattice.
Furthermore, across all models investigated, the dependence of staggered
magnetization on $\Delta$, except very close to $\Delta=1$, is well captured by
SE up to the 12th order. The $\Delta$ dependence of the spin gap closely mirrors
that of staggered magnetization, with most cases being approximately describable
by SE up to the 10th order. However, for the $S=1/2$ honeycomb lattice,
significant deviations from the 10th order SE predictions are observed near
the isotropic Heisenberg limit, underscoring the pivotal role of quantum
fluctuations on the spin gap in its N\'eel state. Moreover, for all models
considered, our results align numerically with the singular behavior of
the spin gap near the isotropic Heisenberg limit as predicted by SWT.

We here delve into why the $\Delta$-dependence of the spin gap for the $S=1/2$ honeycomb lattice case significantly deviates from the SE predictions. The spin gap in this context serves as an indicator of the stability of staggered magnetization in the z-direction within the N\'eel LRO. To analyze this, we consider the ratio of the spin gap to the magnitude of magnetization, normalized to approach unity in the classical limit, defined for the square and honeycomb lattices respectively as:
\begin{align}
	r_{\mathrm{sq}}(S,\Delta)=\frac{\varepsilon_{\mathrm{sq}}(S,\Delta)}{4\Delta m^z_\mathrm{sq}(S,\Delta)}, \
	r_{\mathrm{hon}}(S,\Delta)=\frac{\varepsilon_{\mathrm{hon}}(S,\Delta)}{3\Delta m^z_\mathrm{hon}(S,\Delta)}.
	\label{eq:ratio}
\end{align}
We plot the DMRG results for all four models as a function of $\Delta$ in Fig.\ref{fig:ratio}(a). In all cases, as $\Delta$ approaches 1, increased quantum fluctuations reduce the stability of staggered magnetization, driving the ratio towards zero. Notably, at the same $\Delta$, $r_{\mathrm{hon}}(\frac{1}{2},\Delta)$ is particularly lower than the other three cases, suggesting that the $S=1/2$ honeycomb lattice experiences larger quantum fluctuations relative to its magnetization size, implying a relatively unstable Néel LRO. This observation is consistent with the larger coefficients for higher-order terms in SE. The differences between $r_{\mathrm{hon}}(\frac{1}{2},\Delta)$ and the other three ratios are plotted in Fig.\ref{fig:ratio}(b), revealing the largest quantum fluctuations around $\Delta \approx 1.5$ for the $S=1/2$ honeycomb lattice case, with the $S=1/2$ square lattice next in line, as intuitively expected. Conversely, for $S=1$, this physical quantity appears less dependent on the lattice geometry. The data plotted in Fig.\ref{fig:ratio} are shown in Appendix B.

Additionally, we have tested the application of a recently proposed novel method for numerical calculations in 2D systems, demonstrating its effectiveness. 
The application of SBC has proven to be particularly effective in mapping 2D
models onto 1D periodic chains, significantly simplifying the computational
complexity of our analyses and enabling the precise determination of physical
quantities in the thermodynamic limit. This methodological innovation opens
new avenues for the study of quantum phenomena in complex lattice systems,
providing a robust framework for exploring the effects of lattice geometry and
spin interactions on magnetic order and excitations.

\section*{Acknowledgements}

We thank Ulrike Nitzsche for technical support. This work was supported by Grants-in-Aid for Scientific Research from JSPS (Projects No. JP20H01849, No. JP20K03769, and No. JP21J20604).
M.K. acknowledges support from the JSPS Research Fellowship for Young Scientists.
M.N. acknowledges the research fellow position of the Institute of Industrial Science, The University of Tokyo.
S.N. acknowledges support from SFB 1143 project A05
(project-id 247310070) of the Deutsche Forschungsgemeinschaft.

\appendix

\label{AppA}
\section{Series expansion for spin gap in the $S=1/2$ honeycomb-lattice XXZ Heisenberg model}

\begin{figure}[tbh]
	\centering
	\includegraphics[width=0.7\linewidth]{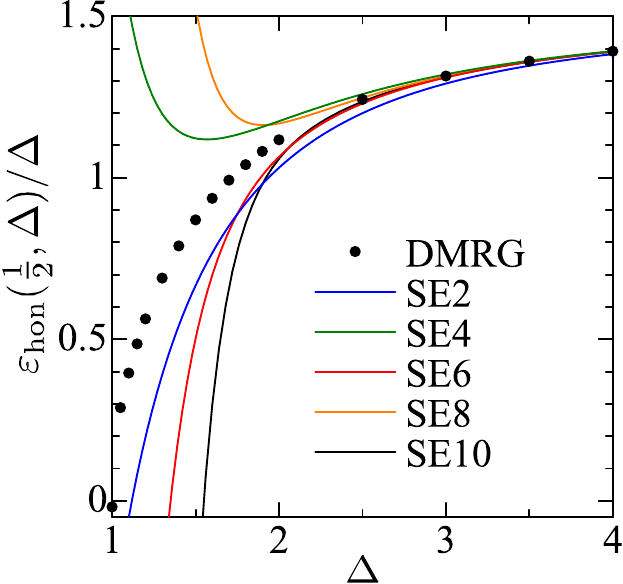}
	\caption{
		$\Delta$-normalized spin gap for the $S=1/2$ honeycomb lattice case
		obtained via DMRG and SE, plotted as a function of $\Delta$. Here,
		the SE results incorporating terms up to the $n$th order are denoted
		as SE$n$.
	}
	\label{fig:se10gaphon}
\end{figure}

Around the classical limit, perturbative expansion up to the 10th order in $1/\Delta$
for the $S=1/2$ honeycomb-lattice XXZ Heisenberg model yields the spin gap as:
\begin{align}
	\nonumber
	\varepsilon_{\mathrm{hon}}(\frac{1}{2},\Delta)&=\frac{3}{2}-\frac{15}{8\Delta^2}+\frac{2.3046875}{\Delta^4}-\frac{7.05102539062}{\Delta^6}\\
	&+\frac{26.3766856347}{\Delta^8}-\frac{111.596182008}{\Delta^{10}}.
	\label{eq:se10gaphon}
\end{align}
We denote the results obtained by considering terms up to the $n$th order as SE$n$
and plot them as a function of $\Delta$ in Fig.~\ref{fig:se10gaphon}, alongside
a comparison with our DMRG results. As indicated by Eq.(\ref{eq:se10gaphon}),
the inclusion of higher-order terms introduces significant oscillations near $\Delta=1$,
due to the increasing coefficients of these terms. Consequently, within the region
$\Delta \lesssim 2$, there is a noticeable deviation from the DMRG results.
To approximate the DMRG findings even closely, it is anticipated that a considerably
higher order of terms must be accounted for, suggesting that the impact of quantum
fluctuations on spin excitations is significant in this Néel phase.

\label{AppB}
\section{DMRG data for staggered magnetization and spin gap}

\begin{table}
	\caption{
		Extrapolated values of staggered magnetization, spin gap, and their ratio
		defined by Eq.~(\ref{eq:ratio}) to the thermodynamic limit as a function of
		$\Delta$ for the $S=1/2$ square-lattice XXZ Heisenberg model obtained
		by our DMRG calculations.
		}
	\begin{center}
		\begin{tabular}{wc{0.6cm}|wc{1.3cm}wc{1.3cm}wc{1.3cm}|wc{1.3cm}wc{1.3cm}wc{1.3cm}}
			\hline
			\hline
			\multirow{2}{*}{$\Delta$} &  \multicolumn{3}{c}{$S=1/2$} & \multicolumn{3}{c}{$S=1$} \\
			& $\varepsilon_{\mathrm{sq}}\!(\frac{1}{2},\Delta)$ & $m^z_{\mathrm{sq}}\!(\frac{1}{2},\Delta)$ & $r_{\mathrm{sq}}\!(\frac{1}{2},\Delta)$ & $\varepsilon_{\mathrm{sq}}\!(1,\Delta)$ & $m^z_{\mathrm{sq}}\!(1,\Delta)$ & $r_{\mathrm{sq}}\!(1,\Delta)$ \\
			\hline
			1.00 & -0.00170 & 0.30651 & -0.00139 & 0.00684  & 0.80170 & 0.00213  \\
			1.05 & 0.46158  & 0.36402 & 0.30191  & 1.11193  & 0.87667 & 0.30199  \\
			1.10 & 0.68102  & 0.38422 & 0.40284  & 1.62512  & 0.89616 & 0.41214  \\
			1.15 & 0.87217  & 0.39752 & 0.47696  & 2.04938  & 0.91139 & 0.48883  \\
			1.20 & 1.04607  & 0.40789 & 0.53429  & 2.42961  & 0.92172 & 0.54915  \\
			1.30 & 1.36273  & 0.42488 & 0.61679  & 3.09541  & 0.93763 & 0.63487  \\
			1.40 & 1.65602  & 0.43681 & 0.67699  & 3.69616  & 0.94862 & 0.69578  \\
			1.50 & 1.93344  & 0.44593 & 0.72262  & 4.25626  & 0.95670 & 0.74149  \\
			1.60 & 2.19954  & 0.45312 & 0.75847  & 4.78831  & 0.96291 & 0.77699  \\
			1.70 & 2.45781  & 0.45891 & 0.78761  & 5.29980  & 0.96780 & 0.80532  \\
			1.80 & 2.70835  & 0.46366 & 0.81128  & 5.79559  & 0.97174 & 0.82836  \\
			1.90 & 2.95351  & 0.46761 & 0.83108  & 6.27901  & 0.97497 & 0.84740  \\
			2.00 & 3.19416  & 0.47094 & 0.84781  & 6.75247  & 0.97765 & 0.86335  \\
			2.50 & 4.34959  & 0.48173 & 0.90292  & 9.02033  & 0.98619 & 0.91467  \\
			3.00 & 5.45464  & 0.48742 & 0.93257  & 11.19124 & 0.99058 & 0.94148  \\
			3.50 & 6.53058  & 0.49081 & 0.95042  & 13.31055 & 0.99315 & 0.95731  \\
			4.00 & 7.58805  & 0.49298 & 0.96200  & 15.39881 & 0.99479 & 0.96747  \\
			\hline                                                
			\hline
		\end{tabular}
	\end{center}
	\label{table:sqdata}
\end{table}

\begin{table}
	\caption{
		Extrapolated values of staggered magnetization, spin gap, and their ratio
		defined by Eq.~(\ref{eq:ratio}) to the thermodynamic limit as a function of
		$\Delta$ for the $S=1/2$ honeycomb-lattice XXZ Heisenberg model obtained
		by our DMRG calculations.
	}
	\begin{center}
		\begin{tabular}{wc{0.6cm}|wc{1.3cm}wc{1.3cm}wc{1.3cm}|wc{1.3cm}wc{1.3cm}wc{1.3cm}}
			\hline
			\hline
			\multirow{2}{*}{$\Delta$} &  \multicolumn{3}{c}{$S=1/2$} & \multicolumn{3}{c}{$S=1$} \\
			& $\varepsilon_{\mathrm{hon}}\!(\frac{1}{2},\Delta)$ & $m^z_{\mathrm{hon}}\!(\frac{1}{2},\Delta)$ & $r_{\mathrm{hon}}\!(\frac{1}{2},\Delta)$ & $\varepsilon_{\mathrm{hon}}\!(1,\Delta)$ & $m^z_{\mathrm{hon}}\!(1,\Delta)$ & $r_{\mathrm{hon}}\!(1,\Delta)$ \\
			\hline
			1.00 & -0.01917 & 0.27637 & -0.02312 & -0.01448 & 0.75925 & -0.00636 \\
			1.05 & 0.30264  & 0.31882 & 0.30135  & 0.77160  & 0.82632 & 0.29644  \\
			1.10 & 0.43473  & 0.34256 & 0.38456  & 1.14549  & 0.85580 & 0.40561  \\
			1.15 & 0.55835  & 0.35957 & 0.45010  & 1.46041  & 0.87559 & 0.48345  \\
			1.20 & 0.67549  & 0.37295 & 0.50312  & 1.73880  & 0.89047 & 0.54241  \\
			1.30 & 0.89595  & 0.39335 & 0.58403  & 2.24251  & 0.91197 & 0.63051  \\
			1.40 & 1.10423  & 0.40856 & 0.64351  & 2.69467  & 0.92701 & 0.69210  \\
			1.50 & 1.30397  & 0.42045 & 0.68919  & 3.11703  & 0.93819 & 0.73831  \\
			1.60 & 1.49774  & 0.43004 & 0.72558  & 3.51818  & 0.94683 & 0.77412  \\
			1.70 & 1.68690  & 0.43792 & 0.75531  & 3.90403  & 0.95368 & 0.80267  \\
			1.80 & 1.87243  & 0.44451 & 0.78007  & 4.27815  & 0.95924 & 0.82592  \\
			1.90 & 2.05486  & 0.45008 & 0.80097  & 4.64300  & 0.96381 & 0.84514  \\
			2.00 & 2.23472  & 0.45484 & 0.81887  & 5.00033  & 0.96764 & 0.86126  \\
			2.50 & 3.10562  & 0.47085 & 0.87944  & 6.71114  & 0.97988 & 0.91319  \\
			3.00 & 3.94431  & 0.47960 & 0.91379  & 8.34704  & 0.98624 & 0.94039  \\
			3.50 & 4.76235  & 0.48494 & 0.93529  & 9.94248  & 0.98997 & 0.95649  \\
			4.00 & 5.56612  & 0.48843 & 0.94967  & 11.51335 & 0.99237 & 0.96683  \\
			\hline                                                
			\hline
		\end{tabular}
	\end{center}
	\label{table:hondata}
\end{table}

For additional context, we present the values of staggered magnetization and spin gap
obtained via the DMRG method at the thermodynamic limit for various values of $\Delta$. Tables~\ref{table:sqdata} and \ref{table:hondata} detail these quantities for the square and honeycomb lattices, respectively. Moreover, the values of the ratio of staggered magnetization to the spin gap, as plotted in Fig.~\ref{fig:ratio} of the main text, are also provided for reference.

\newpage

\bibliography{gaphoneycomb}

\providecommand{\noopsort}[1]{}\providecommand{\singleletter}[1]{#1}%
\begin{thebibliography}{44}%
\makeatletter
\providecommand \@ifxundefined [1]{%
 \@ifx{#1\undefined}
}%
\providecommand \@ifnum [1]{%
 \ifnum #1\expandafter \@firstoftwo
 \else \expandafter \@secondoftwo
 \fi
}%
\providecommand \@ifx [1]{%
 \ifx #1\expandafter \@firstoftwo
 \else \expandafter \@secondoftwo
 \fi
}%
\providecommand \natexlab [1]{#1}%
\providecommand \enquote  [1]{``#1''}%
\providecommand \bibnamefont  [1]{#1}%
\providecommand \bibfnamefont [1]{#1}%
\providecommand \citenamefont [1]{#1}%
\providecommand \href@noop [0]{\@secondoftwo}%
\providecommand \href [0]{\begingroup \@sanitize@url \@href}%
\providecommand \@href[1]{\@@startlink{#1}\@@href}%
\providecommand \@@href[1]{\endgroup#1\@@endlink}%
\providecommand \@sanitize@url [0]{\catcode `\\12\catcode `\$12\catcode
  `\&12\catcode `\#12\catcode `\^12\catcode `\_12\catcode `\%12\relax}%
\providecommand \@@startlink[1]{}%
\providecommand \@@endlink[0]{}%
\providecommand \url  [0]{\begingroup\@sanitize@url \@url }%
\providecommand \@url [1]{\endgroup\@href {#1}{\urlprefix }}%
\providecommand \urlprefix  [0]{URL }%
\providecommand \Eprint [0]{\href }%
\providecommand \doibase [0]{https://doi.org/}%
\providecommand \selectlanguage [0]{\@gobble}%
\providecommand \bibinfo  [0]{\@secondoftwo}%
\providecommand \bibfield  [0]{\@secondoftwo}%
\providecommand \translation [1]{[#1]}%
\providecommand \BibitemOpen [0]{}%
\providecommand \bibitemStop [0]{}%
\providecommand \bibitemNoStop [0]{.\EOS\space}%
\providecommand \EOS [0]{\spacefactor3000\relax}%
\providecommand \BibitemShut  [1]{\csname bibitem#1\endcsname}%
\let\auto@bib@innerbib\@empty
\bibitem [{\citenamefont {Schollw{\"o}ck}\ \emph {et~al.}(2004)\citenamefont
  {Schollw{\"o}ck}, \citenamefont {Richter}, \citenamefont {Farnell},\ and\
  \citenamefont {Bishop}}]{schollwoeck2004}%
  \BibitemOpen
  \bibfield  {author} {\bibinfo {author} {\bibfnamefont {U.}~\bibnamefont
  {Schollw{\"o}ck}}, \bibinfo {author} {\bibfnamefont {J.}~\bibnamefont
  {Richter}}, \bibinfo {author} {\bibfnamefont {D.~J.}\ \bibnamefont
  {Farnell}},\ and\ \bibinfo {author} {\bibfnamefont {R.~F.}\ \bibnamefont
  {Bishop}},\ }\href {https://books.google.de/books?id=ssLvAAAAMAAJ} {\emph
  {\bibinfo {title} {Quantum Magnetism}}},\ \bibinfo {series} {Lecture Notes in
  Physics}, Vol.\ \bibinfo {volume} {645}\ (\bibinfo  {publisher} {Springer},\
  \bibinfo {year} {2004})\BibitemShut {NoStop}%
\bibitem [{\citenamefont {Lacroix}\ \emph {et~al.}(2011)\citenamefont
  {Lacroix}, \citenamefont {Mendels},\ and\ \citenamefont
  {Mila}}]{lacroix2011}%
  \BibitemOpen
  \bibfield  {author} {\bibinfo {author} {\bibfnamefont {C.}~\bibnamefont
  {Lacroix}}, \bibinfo {author} {\bibfnamefont {P.}~\bibnamefont {Mendels}},\
  and\ \bibinfo {author} {\bibfnamefont {F.}~\bibnamefont {Mila}},\ }\href@noop
  {} {\emph {\bibinfo {title} {Introduction to frustrated magnetism: materials,
  experiments, theory}}},\ Vol.\ \bibinfo {volume} {164}\ (\bibinfo
  {publisher} {Springer Science \& Business Media},\ \bibinfo {year}
  {2011})\BibitemShut {NoStop}%
\bibitem [{\citenamefont {Sachdev}(2011)}]{sachdev2011}%
  \BibitemOpen
  \bibfield  {author} {\bibinfo {author} {\bibfnamefont {S.}~\bibnamefont
  {Sachdev}},\ }\href {https://books.google.de/books?id=OqcEmQEACAAJ} {\emph
  {\bibinfo {title} {Quantum Phase Transitions}}}\ (\bibinfo  {publisher}
  {Cambridge University Press},\ \bibinfo {year} {2011})\BibitemShut {NoStop}%
\bibitem [{\citenamefont {Heisenberg}(1928)}]{Heisenberg1928}%
  \BibitemOpen
  \bibfield  {author} {\bibinfo {author} {\bibfnamefont {W.}~\bibnamefont
  {Heisenberg}},\ }\bibfield  {title} {\bibinfo {title} {Zur theorie des
  ferromagnetismus},\ }\href {https://doi.org/10.1007/BF01328601} {\bibfield
  {journal} {\bibinfo  {journal} {Z. Phys.}\ }\textbf {\bibinfo {volume}
  {49}},\ \bibinfo {pages} {619} (\bibinfo {year} {1928})}\BibitemShut
  {NoStop}%
\bibitem [{\citenamefont {Lines}(1963)}]{Lines1963}%
  \BibitemOpen
  \bibfield  {author} {\bibinfo {author} {\bibfnamefont {M.~E.}\ \bibnamefont
  {Lines}},\ }\bibfield  {title} {\bibinfo {title} {Magnetic properties of
  co${\mathrm{cl}}_{2}$ and ni${\mathrm{cl}}_{2}$},\ }\href
  {https://doi.org/10.1103/PhysRev.131.546} {\bibfield  {journal} {\bibinfo
  {journal} {Phys. Rev.}\ }\textbf {\bibinfo {volume} {131}},\ \bibinfo {pages}
  {546} (\bibinfo {year} {1963})}\BibitemShut {NoStop}%
\bibitem [{\citenamefont {Achiwa}(1969)}]{Achiwa1969}%
  \BibitemOpen
  \bibfield  {author} {\bibinfo {author} {\bibfnamefont {N.}~\bibnamefont
  {Achiwa}},\ }\bibfield  {title} {\bibinfo {title} {Linear antiferromagnetic
  chains in hexagonal abcl3-type compounds (a; cs, or rb, b; cu, ni, co, or
  fe)},\ }\href {https://doi.org/10.1143/JPSJ.27.561} {\bibfield  {journal}
  {\bibinfo  {journal} {Journal of the Physical Society of Japan}\ }\textbf
  {\bibinfo {volume} {27}},\ \bibinfo {pages} {561} (\bibinfo {year} {1969})},\
  \Eprint {https://arxiv.org/abs/https://doi.org/10.1143/JPSJ.27.561}
  {https://doi.org/10.1143/JPSJ.27.561} \BibitemShut {NoStop}%
\bibitem [{\citenamefont {Savary}\ and\ \citenamefont
  {Balents}(2016)}]{savary2016}%
  \BibitemOpen
  \bibfield  {author} {\bibinfo {author} {\bibfnamefont {L.}~\bibnamefont
  {Savary}}\ and\ \bibinfo {author} {\bibfnamefont {L.}~\bibnamefont
  {Balents}},\ }\bibfield  {title} {\bibinfo {title} {Quantum spin liquids: a
  review},\ }\href@noop {} {\bibfield  {journal} {\bibinfo  {journal} {Reports
  on Progress in Physics}\ }\textbf {\bibinfo {volume} {80}},\ \bibinfo {pages}
  {016502} (\bibinfo {year} {2016})}\BibitemShut {NoStop}%
\bibitem [{\citenamefont {Zhou}\ \emph
  {et~al.}(2017{\natexlab{a}})\citenamefont {Zhou}, \citenamefont {Kanoda},\
  and\ \citenamefont {Ng}}]{zhou2017}%
  \BibitemOpen
  \bibfield  {author} {\bibinfo {author} {\bibfnamefont {Y.}~\bibnamefont
  {Zhou}}, \bibinfo {author} {\bibfnamefont {K.}~\bibnamefont {Kanoda}},\ and\
  \bibinfo {author} {\bibfnamefont {T.-K.}\ \bibnamefont {Ng}},\ }\bibfield
  {title} {\bibinfo {title} {Quantum spin liquid states},\ }\href@noop {}
  {\bibfield  {journal} {\bibinfo  {journal} {Reviews of Modern Physics}\
  }\textbf {\bibinfo {volume} {89}},\ \bibinfo {pages} {025003} (\bibinfo
  {year} {2017}{\natexlab{a}})}\BibitemShut {NoStop}%
\bibitem [{\citenamefont {Schrieffer}\ and\ \citenamefont
  {Brooks}(2007)}]{schrieffer2007}%
  \BibitemOpen
  \bibfield  {author} {\bibinfo {author} {\bibfnamefont {J.~R.}\ \bibnamefont
  {Schrieffer}}\ and\ \bibinfo {author} {\bibfnamefont {J.~S.}\ \bibnamefont
  {Brooks}},\ }\href@noop {} {\emph {\bibinfo {title} {Handbook of
  high-temperature superconductivity: theory and experiment}}}\ (\bibinfo
  {publisher} {Springer},\ \bibinfo {year} {2007})\BibitemShut {NoStop}%
\bibitem [{\citenamefont {Bernevig}\ \emph {et~al.}(2006)\citenamefont
  {Bernevig}, \citenamefont {Hughes},\ and\ \citenamefont
  {Zhang}}]{bernevig2006}%
  \BibitemOpen
  \bibfield  {author} {\bibinfo {author} {\bibfnamefont {B.~A.}\ \bibnamefont
  {Bernevig}}, \bibinfo {author} {\bibfnamefont {T.~L.}\ \bibnamefont
  {Hughes}},\ and\ \bibinfo {author} {\bibfnamefont {S.-C.}\ \bibnamefont
  {Zhang}},\ }\bibfield  {title} {\bibinfo {title} {Quantum spin hall effect
  and topological phase transition in hgte quantum wells},\ }\href@noop {}
  {\bibfield  {journal} {\bibinfo  {journal} {science}\ }\textbf {\bibinfo
  {volume} {314}},\ \bibinfo {pages} {1757} (\bibinfo {year}
  {2006})}\BibitemShut {NoStop}%
\bibitem [{\citenamefont {Zhou}\ \emph
  {et~al.}(2017{\natexlab{b}})\citenamefont {Zhou}, \citenamefont {Kanoda},\
  and\ \citenamefont {Ng}}]{Yi2017}%
  \BibitemOpen
  \bibfield  {author} {\bibinfo {author} {\bibfnamefont {Y.}~\bibnamefont
  {Zhou}}, \bibinfo {author} {\bibfnamefont {K.}~\bibnamefont {Kanoda}},\ and\
  \bibinfo {author} {\bibfnamefont {T.-K.}\ \bibnamefont {Ng}},\ }\bibfield
  {title} {\bibinfo {title} {Quantum spin liquid states},\ }\href
  {https://doi.org/10.1103/RevModPhys.89.025003} {\bibfield  {journal}
  {\bibinfo  {journal} {Rev. Mod. Phys.}\ }\textbf {\bibinfo {volume} {89}},\
  \bibinfo {pages} {025003} (\bibinfo {year} {2017}{\natexlab{b}})}\BibitemShut
  {NoStop}%
\bibitem [{\citenamefont {Kadosawa}\ \emph {et~al.}(2022)\citenamefont
  {Kadosawa}, \citenamefont {Nakamura}, \citenamefont {Ohta},\ and\
  \citenamefont {Nishimoto}}]{Kadosawa2022}%
  \BibitemOpen
  \bibfield  {author} {\bibinfo {author} {\bibfnamefont {M.}~\bibnamefont
  {Kadosawa}}, \bibinfo {author} {\bibfnamefont {M.}~\bibnamefont {Nakamura}},
  \bibinfo {author} {\bibfnamefont {Y.}~\bibnamefont {Ohta}},\ and\ \bibinfo
  {author} {\bibfnamefont {S.}~\bibnamefont {Nishimoto}},\ }\bibfield  {title}
  {\bibinfo {title} {One-dimensional projection of two-dimensional systems
  using spiral boundary conditions},\ }\href {https://arxiv.org/abs/2205.15775}
  {\bibfield  {journal} {\bibinfo  {journal} {arXiv:2205.15775}\ } (\bibinfo
  {year} {2022})}\BibitemShut {NoStop}%
\bibitem [{\citenamefont {Kadosawa}\ \emph
  {et~al.}(2023{\natexlab{a}})\citenamefont {Kadosawa}, \citenamefont
  {Nakamura}, \citenamefont {Ohta},\ and\ \citenamefont
  {Nishimoto}}]{Kadosawa2023-1}%
  \BibitemOpen
  \bibfield  {author} {\bibinfo {author} {\bibfnamefont {M.}~\bibnamefont
  {Kadosawa}}, \bibinfo {author} {\bibfnamefont {M.}~\bibnamefont {Nakamura}},
  \bibinfo {author} {\bibfnamefont {Y.}~\bibnamefont {Ohta}},\ and\ \bibinfo
  {author} {\bibfnamefont {S.}~\bibnamefont {Nishimoto}},\ }\bibfield  {title}
  {\bibinfo {title} {Study of staggered magnetization in the spin-s
  square-lattice heisenberg model using spiral boundary conditions},\ }\href
  {https://doi.org/10.7566/JPSJ.92.023701} {\bibfield  {journal} {\bibinfo
  {journal} {Journal of the Physical Society of Japan}\ }\textbf {\bibinfo
  {volume} {92}},\ \bibinfo {pages} {023701} (\bibinfo {year}
  {2023}{\natexlab{a}})}\BibitemShut {NoStop}%
\bibitem [{\citenamefont {Kubo}\ and\ \citenamefont {Kishi}(1988)}]{Kubo1988}%
  \BibitemOpen
  \bibfield  {author} {\bibinfo {author} {\bibfnamefont {K.}~\bibnamefont
  {Kubo}}\ and\ \bibinfo {author} {\bibfnamefont {T.}~\bibnamefont {Kishi}},\
  }\bibfield  {title} {\bibinfo {title} {Existence of long-range order in the
  $\mathrm{XXZ}$ model},\ }\href {https://doi.org/10.1103/PhysRevLett.61.2585}
  {\bibfield  {journal} {\bibinfo  {journal} {Phys. Rev. Lett.}\ }\textbf
  {\bibinfo {volume} {61}},\ \bibinfo {pages} {2585} (\bibinfo {year}
  {1988})}\BibitemShut {NoStop}%
\bibitem [{\citenamefont {Weihong}\ \emph
  {et~al.}(1991{\natexlab{a}})\citenamefont {Weihong}, \citenamefont {Oitmaa},\
  and\ \citenamefont {Hamer}}]{Weihong1991-2}%
  \BibitemOpen
  \bibfield  {author} {\bibinfo {author} {\bibfnamefont {Z.}~\bibnamefont
  {Weihong}}, \bibinfo {author} {\bibfnamefont {J.}~\bibnamefont {Oitmaa}},\
  and\ \bibinfo {author} {\bibfnamefont {C.~J.}\ \bibnamefont {Hamer}},\
  }\bibfield  {title} {\bibinfo {title} {Second-order spin-wave results for the
  quantum xxz and xy models with anisotropy},\ }\href
  {https://doi.org/10.1103/PhysRevB.44.11869} {\bibfield  {journal} {\bibinfo
  {journal} {Phys. Rev. B}\ }\textbf {\bibinfo {volume} {44}},\ \bibinfo
  {pages} {11869} (\bibinfo {year} {1991}{\natexlab{a}})}\BibitemShut {NoStop}%
\bibitem [{\citenamefont {Viswanath}\ \emph {et~al.}(1994)\citenamefont
  {Viswanath}, \citenamefont {Zhang}, \citenamefont {Stolze},\ and\
  \citenamefont {M\"uller}}]{Viswanath1994}%
  \BibitemOpen
  \bibfield  {author} {\bibinfo {author} {\bibfnamefont {V.~S.}\ \bibnamefont
  {Viswanath}}, \bibinfo {author} {\bibfnamefont {S.}~\bibnamefont {Zhang}},
  \bibinfo {author} {\bibfnamefont {J.}~\bibnamefont {Stolze}},\ and\ \bibinfo
  {author} {\bibfnamefont {G.}~\bibnamefont {M\"uller}},\ }\bibfield  {title}
  {\bibinfo {title} {Ordering and fluctuations in the ground state of the
  one-dimensional and two-dimensional s=1/2 xxz antiferromagnets: A study of
  dynamical properties based on the recursion method},\ }\href
  {https://doi.org/10.1103/PhysRevB.49.9702} {\bibfield  {journal} {\bibinfo
  {journal} {Phys. Rev. B}\ }\textbf {\bibinfo {volume} {49}},\ \bibinfo
  {pages} {9702} (\bibinfo {year} {1994})}\BibitemShut {NoStop}%
\bibitem [{\citenamefont {Yunoki}(2002)}]{Yunoki2002}%
  \BibitemOpen
  \bibfield  {author} {\bibinfo {author} {\bibfnamefont {S.}~\bibnamefont
  {Yunoki}},\ }\bibfield  {title} {\bibinfo {title} {Numerical study of the
  spin-flop transition in anisotropic spin-$\frac{1}{2}$ antiferromagnets},\
  }\href {https://doi.org/10.1103/PhysRevB.65.092402} {\bibfield  {journal}
  {\bibinfo  {journal} {Phys. Rev. B}\ }\textbf {\bibinfo {volume} {65}},\
  \bibinfo {pages} {092402} (\bibinfo {year} {2002})}\BibitemShut {NoStop}%
\bibitem [{\citenamefont {Braiorr-Orrs}\ \emph {et~al.}(2019)\citenamefont
  {Braiorr-Orrs}, \citenamefont {Weyrauch},\ and\ \citenamefont
  {Rakov}}]{Braiorr-Orrs2019}%
  \BibitemOpen
  \bibfield  {author} {\bibinfo {author} {\bibfnamefont {B.}~\bibnamefont
  {Braiorr-Orrs}}, \bibinfo {author} {\bibfnamefont {M.}~\bibnamefont
  {Weyrauch}},\ and\ \bibinfo {author} {\bibfnamefont {M.~V.}\ \bibnamefont
  {Rakov}},\ }\bibfield  {title} {\bibinfo {title} {Numerical studies of
  entanglement properties in one- and two-dimensional quantum ising and xxz
  models},\ }\href {https://doi.org/10.15407/ujpe61.07.0613} {\bibfield
  {journal} {\bibinfo  {journal} {Ukrainian Journal of Physics}\ }\textbf
  {\bibinfo {volume} {61}},\ \bibinfo {pages} {613} (\bibinfo {year}
  {2019})}\BibitemShut {NoStop}%
\bibitem [{\citenamefont {Bishop}\ and\ \citenamefont {Li}(2016)}]{Bishop2016}%
  \BibitemOpen
  \bibfield  {author} {\bibinfo {author} {\bibfnamefont {R.}~\bibnamefont
  {Bishop}}\ and\ \bibinfo {author} {\bibfnamefont {P.}~\bibnamefont {Li}},\
  }\bibfield  {title} {\bibinfo {title} {Large-s expansions for the low-energy
  parameters of the honeycomb-lattice heisenberg antiferromagnet with spin
  quantum number s},\ }\href
  {https://doi.org/https://doi.org/10.1016/j.jmmm.2016.01.101} {\bibfield
  {journal} {\bibinfo  {journal} {Journal of Magnetism and Magnetic Materials}\
  }\textbf {\bibinfo {volume} {407}},\ \bibinfo {pages} {348} (\bibinfo {year}
  {2016})}\BibitemShut {NoStop}%
\bibitem [{\citenamefont {Affleck}\ \emph {et~al.}(1988)\citenamefont
  {Affleck}, \citenamefont {Kennedy}, \citenamefont {Lieb},\ and\ \citenamefont
  {Tasaki}}]{Affleck1988}%
  \BibitemOpen
  \bibfield  {author} {\bibinfo {author} {\bibfnamefont {I.}~\bibnamefont
  {Affleck}}, \bibinfo {author} {\bibfnamefont {T.}~\bibnamefont {Kennedy}},
  \bibinfo {author} {\bibfnamefont {E.~H.}\ \bibnamefont {Lieb}},\ and\
  \bibinfo {author} {\bibfnamefont {H.}~\bibnamefont {Tasaki}},\ }\bibfield
  {title} {\bibinfo {title} {Valence bond ground states in isotropic quantum
  antiferromagnets},\ }\href@noop {} {\bibfield  {journal} {\bibinfo  {journal}
  {Communications in Mathematical Physics}\ }\textbf {\bibinfo {volume}
  {115}},\ \bibinfo {pages} {477} (\bibinfo {year} {1988})}\BibitemShut
  {NoStop}%
\bibitem [{\citenamefont {Wojtkiewicz}\ \emph {et~al.}(2023)\citenamefont
  {Wojtkiewicz}, \citenamefont {Wohlfeld},\ and\ \citenamefont
  {Ole\ifmmode~\acute{s}\else \'{s}\fi{}}}]{Wojtkiewicz2023}%
  \BibitemOpen
  \bibfield  {author} {\bibinfo {author} {\bibfnamefont {J.}~\bibnamefont
  {Wojtkiewicz}}, \bibinfo {author} {\bibfnamefont {K.}~\bibnamefont
  {Wohlfeld}},\ and\ \bibinfo {author} {\bibfnamefont {A.~M.}\ \bibnamefont
  {Ole\ifmmode~\acute{s}\else \'{s}\fi{}}},\ }\bibfield  {title} {\bibinfo
  {title} {Long-range order in the xy model on the honeycomb lattice},\ }\href
  {https://doi.org/10.1103/PhysRevB.107.064409} {\bibfield  {journal} {\bibinfo
   {journal} {Phys. Rev. B}\ }\textbf {\bibinfo {volume} {107}},\ \bibinfo
  {pages} {064409} (\bibinfo {year} {2023})}\BibitemShut {NoStop}%
\bibitem [{\citenamefont {Nakamura}\ \emph {et~al.}(2021)\citenamefont
  {Nakamura}, \citenamefont {Masuda},\ and\ \citenamefont
  {Nishimoto}}]{Nakamura2021}%
  \BibitemOpen
  \bibfield  {author} {\bibinfo {author} {\bibfnamefont {M.}~\bibnamefont
  {Nakamura}}, \bibinfo {author} {\bibfnamefont {S.}~\bibnamefont {Masuda}},\
  and\ \bibinfo {author} {\bibfnamefont {S.}~\bibnamefont {Nishimoto}},\
  }\bibfield  {title} {\bibinfo {title} {Characterization of topological
  insulators based on the electronic polarization with spiral boundary
  conditions},\ }\href {https://doi.org/10.1103/PhysRevB.104.L121114}
  {\bibfield  {journal} {\bibinfo  {journal} {Phys. Rev. B}\ }\textbf {\bibinfo
  {volume} {104}},\ \bibinfo {pages} {L121114} (\bibinfo {year}
  {2021})}\BibitemShut {NoStop}%
\bibitem [{\citenamefont {White}(1992)}]{White1992}%
  \BibitemOpen
  \bibfield  {author} {\bibinfo {author} {\bibfnamefont {S.~R.}\ \bibnamefont
  {White}},\ }\bibfield  {title} {\bibinfo {title} {Density matrix formulation
  for quantum renormalization groups},\ }\href
  {https://doi.org/10.1103/PhysRevLett.69.2863} {\bibfield  {journal} {\bibinfo
   {journal} {Phys. Rev. Lett.}\ }\textbf {\bibinfo {volume} {69}},\ \bibinfo
  {pages} {2863} (\bibinfo {year} {1992})}\BibitemShut {NoStop}%
\bibitem [{\citenamefont {Kadosawa}\ \emph
  {et~al.}(2023{\natexlab{b}})\citenamefont {Kadosawa}, \citenamefont
  {Nakamura}, \citenamefont {Ohta},\ and\ \citenamefont
  {Nishimoto}}]{Kadosawa2023-2}%
  \BibitemOpen
  \bibfield  {author} {\bibinfo {author} {\bibfnamefont {M.}~\bibnamefont
  {Kadosawa}}, \bibinfo {author} {\bibfnamefont {M.}~\bibnamefont {Nakamura}},
  \bibinfo {author} {\bibfnamefont {Y.}~\bibnamefont {Ohta}},\ and\ \bibinfo
  {author} {\bibfnamefont {S.}~\bibnamefont {Nishimoto}},\ }\bibfield  {title}
  {\bibinfo {title} {Phase diagram of the kitaev-heisenberg model using various
  finite-size clusters},\ }\href {https://doi.org/10.7566/JPSJ.92.055001}
  {\bibfield  {journal} {\bibinfo  {journal} {Journal of the Physical Society
  of Japan}\ }\textbf {\bibinfo {volume} {92}},\ \bibinfo {pages} {055001}
  (\bibinfo {year} {2023}{\natexlab{b}})}\BibitemShut {NoStop}%
\bibitem [{\citenamefont {Singh}(1989)}]{Singh1989}%
  \BibitemOpen
  \bibfield  {author} {\bibinfo {author} {\bibfnamefont {R.~R.~P.}\
  \bibnamefont {Singh}},\ }\bibfield  {title} {\bibinfo {title} {Thermodynamic
  parameters of the $t=0$, spin-1/2 square-lattice heisenberg
  antiferromagnet},\ }\href {https://doi.org/10.1103/PhysRevB.39.9760}
  {\bibfield  {journal} {\bibinfo  {journal} {Phys. Rev. B}\ }\textbf {\bibinfo
  {volume} {39}},\ \bibinfo {pages} {9760} (\bibinfo {year}
  {1989})}\BibitemShut {NoStop}%
\bibitem [{\citenamefont {Singh}(1990)}]{Singh1990}%
  \BibitemOpen
  \bibfield  {author} {\bibinfo {author} {\bibfnamefont {R.~R.~P.}\
  \bibnamefont {Singh}},\ }\bibfield  {title} {\bibinfo {title} {Quantum
  renormalizations in the spin-1 heisenberg antiferromagnet on the square
  lattice},\ }\href {https://doi.org/10.1103/PhysRevB.41.4873} {\bibfield
  {journal} {\bibinfo  {journal} {Phys. Rev. B}\ }\textbf {\bibinfo {volume}
  {41}},\ \bibinfo {pages} {4873} (\bibinfo {year} {1990})}\BibitemShut
  {NoStop}%
\bibitem [{\citenamefont {Weihong}\ \emph
  {et~al.}(1991{\natexlab{b}})\citenamefont {Weihong}, \citenamefont {Oitmaa},\
  and\ \citenamefont {Hamer}}]{Weihong1991-1}%
  \BibitemOpen
  \bibfield  {author} {\bibinfo {author} {\bibfnamefont {Z.}~\bibnamefont
  {Weihong}}, \bibinfo {author} {\bibfnamefont {J.}~\bibnamefont {Oitmaa}},\
  and\ \bibinfo {author} {\bibfnamefont {C.~J.}\ \bibnamefont {Hamer}},\
  }\bibfield  {title} {\bibinfo {title} {Square-lattice heisenberg
  antiferromagnet at $t=0$},\ }\href {https://doi.org/10.1103/PhysRevB.43.8321}
  {\bibfield  {journal} {\bibinfo  {journal} {Phys. Rev. B}\ }\textbf {\bibinfo
  {volume} {43}},\ \bibinfo {pages} {8321} (\bibinfo {year}
  {1991}{\natexlab{b}})}\BibitemShut {NoStop}%
\bibitem [{\citenamefont {Bishop}\ \emph {et~al.}(2017)\citenamefont {Bishop},
  \citenamefont {Li}, \citenamefont {Zinke}, \citenamefont {Darradi},
  \citenamefont {Richter}, \citenamefont {Farnell},\ and\ \citenamefont
  {Schulenburg}}]{BISHOP2017}%
  \BibitemOpen
  \bibfield  {author} {\bibinfo {author} {\bibfnamefont {R.}~\bibnamefont
  {Bishop}}, \bibinfo {author} {\bibfnamefont {P.}~\bibnamefont {Li}}, \bibinfo
  {author} {\bibfnamefont {R.}~\bibnamefont {Zinke}}, \bibinfo {author}
  {\bibfnamefont {R.}~\bibnamefont {Darradi}}, \bibinfo {author} {\bibfnamefont
  {J.}~\bibnamefont {Richter}}, \bibinfo {author} {\bibfnamefont
  {D.}~\bibnamefont {Farnell}},\ and\ \bibinfo {author} {\bibfnamefont
  {J.}~\bibnamefont {Schulenburg}},\ }\bibfield  {title} {\bibinfo {title} {The
  spin-half xxz antiferromagnet on the square lattice revisited: A high-order
  coupled cluster treatment},\ }\href
  {https://doi.org/https://doi.org/10.1016/j.jmmm.2016.11.043} {\bibfield
  {journal} {\bibinfo  {journal} {Journal of Magnetism and Magnetic Materials}\
  }\textbf {\bibinfo {volume} {428}},\ \bibinfo {pages} {178} (\bibinfo {year}
  {2017})}\BibitemShut {NoStop}%
\bibitem [{\citenamefont {White}\ and\ \citenamefont
  {Chernyshev}(2007)}]{White2007}%
  \BibitemOpen
  \bibfield  {author} {\bibinfo {author} {\bibfnamefont {S.~R.}\ \bibnamefont
  {White}}\ and\ \bibinfo {author} {\bibfnamefont {A.~L.}\ \bibnamefont
  {Chernyshev}},\ }\bibfield  {title} {\bibinfo {title} {Ne\'el order in square
  and triangular lattice heisenberg models},\ }\href
  {https://doi.org/10.1103/PhysRevLett.99.127004} {\bibfield  {journal}
  {\bibinfo  {journal} {Phys. Rev. Lett.}\ }\textbf {\bibinfo {volume} {99}},\
  \bibinfo {pages} {127004} (\bibinfo {year} {2007})}\BibitemShut {NoStop}%
\bibitem [{\citenamefont {Sandvik}\ and\ \citenamefont
  {Evertz}(2010)}]{Sandvik2010}%
  \BibitemOpen
  \bibfield  {author} {\bibinfo {author} {\bibfnamefont {A.~W.}\ \bibnamefont
  {Sandvik}}\ and\ \bibinfo {author} {\bibfnamefont {H.~G.}\ \bibnamefont
  {Evertz}},\ }\bibfield  {title} {\bibinfo {title} {Loop updates for
  variational and projector quantum monte carlo simulations in the valence-bond
  basis},\ }\href {https://doi.org/10.1103/PhysRevB.82.024407} {\bibfield
  {journal} {\bibinfo  {journal} {Phys. Rev. B}\ }\textbf {\bibinfo {volume}
  {82}},\ \bibinfo {pages} {024407} (\bibinfo {year} {2010})}\BibitemShut
  {NoStop}%
\bibitem [{\citenamefont {Niesen}\ and\ \citenamefont
  {Corboz}(2017)}]{Niesen2017}%
  \BibitemOpen
  \bibfield  {author} {\bibinfo {author} {\bibfnamefont {I.}~\bibnamefont
  {Niesen}}\ and\ \bibinfo {author} {\bibfnamefont {P.}~\bibnamefont
  {Corboz}},\ }\bibfield  {title} {\bibinfo {title} {Emergent haldane phase in
  the $s=1$ bilinear-biquadratic heisenberg model on the square lattice},\
  }\href {https://doi.org/10.1103/PhysRevB.95.180404} {\bibfield  {journal}
  {\bibinfo  {journal} {Phys. Rev. B}\ }\textbf {\bibinfo {volume} {95}},\
  \bibinfo {pages} {180404} (\bibinfo {year} {2017})}\BibitemShut {NoStop}%
\bibitem [{\citenamefont {Hamer}\ \emph {et~al.}(1992)\citenamefont {Hamer},
  \citenamefont {Weihong},\ and\ \citenamefont {Arndt}}]{Hamer1992}%
  \BibitemOpen
  \bibfield  {author} {\bibinfo {author} {\bibfnamefont {C.~J.}\ \bibnamefont
  {Hamer}}, \bibinfo {author} {\bibfnamefont {Z.}~\bibnamefont {Weihong}},\
  and\ \bibinfo {author} {\bibfnamefont {P.}~\bibnamefont {Arndt}},\ }\bibfield
   {title} {\bibinfo {title} {Third-order spin-wave theory for the heisenberg
  antiferromagnet},\ }\href {https://doi.org/10.1103/PhysRevB.46.6276}
  {\bibfield  {journal} {\bibinfo  {journal} {Phys. Rev. B}\ }\textbf {\bibinfo
  {volume} {46}},\ \bibinfo {pages} {6276} (\bibinfo {year}
  {1992})}\BibitemShut {NoStop}%
\bibitem [{\citenamefont {Igarashi}(1992)}]{Igarashi1992}%
  \BibitemOpen
  \bibfield  {author} {\bibinfo {author} {\bibfnamefont {J.-i.}\ \bibnamefont
  {Igarashi}},\ }\bibfield  {title} {\bibinfo {title} {1/s expansion for
  thermodynamic quantities in a two-dimensional heisenberg antiferromagnet at
  zero temperature},\ }\href {https://doi.org/10.1103/PhysRevB.46.10763}
  {\bibfield  {journal} {\bibinfo  {journal} {Phys. Rev. B}\ }\textbf {\bibinfo
  {volume} {46}},\ \bibinfo {pages} {10763} (\bibinfo {year}
  {1992})}\BibitemShut {NoStop}%
\bibitem [{\citenamefont {Canali}\ and\ \citenamefont
  {Wallin}(1993)}]{Canali1993}%
  \BibitemOpen
  \bibfield  {author} {\bibinfo {author} {\bibfnamefont {C.~M.}\ \bibnamefont
  {Canali}}\ and\ \bibinfo {author} {\bibfnamefont {M.}~\bibnamefont
  {Wallin}},\ }\bibfield  {title} {\bibinfo {title} {Spin-spin correlation
  functions for the square-lattice heisenberg antiferromagnet at zero
  temperature},\ }\href {https://doi.org/10.1103/PhysRevB.48.3264} {\bibfield
  {journal} {\bibinfo  {journal} {Phys. Rev. B}\ }\textbf {\bibinfo {volume}
  {48}},\ \bibinfo {pages} {3264} (\bibinfo {year} {1993})}\BibitemShut
  {NoStop}%
\bibitem [{\citenamefont {Davis}(1960)}]{Davis1960}%
  \BibitemOpen
  \bibfield  {author} {\bibinfo {author} {\bibfnamefont {H.~L.}\ \bibnamefont
  {Davis}},\ }\bibfield  {title} {\bibinfo {title} {New method for treating the
  antiferromagnetic ground state},\ }\href
  {https://doi.org/10.1103/PhysRev.120.789} {\bibfield  {journal} {\bibinfo
  {journal} {Phys. Rev.}\ }\textbf {\bibinfo {volume} {120}},\ \bibinfo {pages}
  {789} (\bibinfo {year} {1960})}\BibitemShut {NoStop}%
\bibitem [{\citenamefont {Huse}(1988)}]{Huse1988}%
  \BibitemOpen
  \bibfield  {author} {\bibinfo {author} {\bibfnamefont {D.~A.}\ \bibnamefont
  {Huse}},\ }\bibfield  {title} {\bibinfo {title} {Ground-state staggered
  magnetization of two-dimensional quantum heisenberg antiferromagnets},\
  }\href {https://doi.org/10.1103/PhysRevB.37.2380} {\bibfield  {journal}
  {\bibinfo  {journal} {Phys. Rev. B}\ }\textbf {\bibinfo {volume} {37}},\
  \bibinfo {pages} {2380} (\bibinfo {year} {1988})}\BibitemShut {NoStop}%
\bibitem [{\citenamefont {Parrinello}\ and\ \citenamefont
  {Arai}(1974)}]{Parrinello1974}%
  \BibitemOpen
  \bibfield  {author} {\bibinfo {author} {\bibfnamefont {M.}~\bibnamefont
  {Parrinello}}\ and\ \bibinfo {author} {\bibfnamefont {T.}~\bibnamefont
  {Arai}},\ }\bibfield  {title} {\bibinfo {title} {Infinite-order cumulant
  expansion for spins},\ }\href {https://doi.org/10.1103/PhysRevB.10.265}
  {\bibfield  {journal} {\bibinfo  {journal} {Phys. Rev. B}\ }\textbf {\bibinfo
  {volume} {10}},\ \bibinfo {pages} {265} (\bibinfo {year} {1974})}\BibitemShut
  {NoStop}%
\bibitem [{\citenamefont {Oitmaa}\ \emph {et~al.}(1992)\citenamefont {Oitmaa},
  \citenamefont {Hamer},\ and\ \citenamefont {Weihong}}]{Oitmaa1992}%
  \BibitemOpen
  \bibfield  {author} {\bibinfo {author} {\bibfnamefont {J.}~\bibnamefont
  {Oitmaa}}, \bibinfo {author} {\bibfnamefont {C.~J.}\ \bibnamefont {Hamer}},\
  and\ \bibinfo {author} {\bibfnamefont {Z.}~\bibnamefont {Weihong}},\
  }\bibfield  {title} {\bibinfo {title} {Quantum magnets on the honeycomb and
  triangular lattices at t=0},\ }\href
  {https://doi.org/10.1103/PhysRevB.45.9834} {\bibfield  {journal} {\bibinfo
  {journal} {Phys. Rev. B}\ }\textbf {\bibinfo {volume} {45}},\ \bibinfo
  {pages} {9834} (\bibinfo {year} {1992})}\BibitemShut {NoStop}%
\bibitem [{\citenamefont {Ganesh}\ \emph {et~al.}(2013)\citenamefont {Ganesh},
  \citenamefont {van~den Brink},\ and\ \citenamefont {Nishimoto}}]{Ganesh2013}%
  \BibitemOpen
  \bibfield  {author} {\bibinfo {author} {\bibfnamefont {R.}~\bibnamefont
  {Ganesh}}, \bibinfo {author} {\bibfnamefont {J.}~\bibnamefont {van~den
  Brink}},\ and\ \bibinfo {author} {\bibfnamefont {S.}~\bibnamefont
  {Nishimoto}},\ }\bibfield  {title} {\bibinfo {title} {Deconfined criticality
  in the frustrated heisenberg honeycomb antiferromagnet},\ }\href
  {https://doi.org/10.1103/PhysRevLett.110.127203} {\bibfield  {journal}
  {\bibinfo  {journal} {Phys. Rev. Lett.}\ }\textbf {\bibinfo {volume} {110}},\
  \bibinfo {pages} {127203} (\bibinfo {year} {2013})}\BibitemShut {NoStop}%
\bibitem [{\citenamefont {Zhu}\ \emph {et~al.}(2013)\citenamefont {Zhu},
  \citenamefont {Huse},\ and\ \citenamefont {White}}]{Zhu2013}%
  \BibitemOpen
  \bibfield  {author} {\bibinfo {author} {\bibfnamefont {Z.}~\bibnamefont
  {Zhu}}, \bibinfo {author} {\bibfnamefont {D.~A.}\ \bibnamefont {Huse}},\ and\
  \bibinfo {author} {\bibfnamefont {S.~R.}\ \bibnamefont {White}},\ }\bibfield
  {title} {\bibinfo {title} {Weak plaquette valence bond order in the
  $s\mathbf{=}1/2$ honeycomb ${J}_{1}\mathbf{\ensuremath{-}}{J}_{2}$ heisenberg
  model},\ }\href {https://doi.org/10.1103/PhysRevLett.110.127205} {\bibfield
  {journal} {\bibinfo  {journal} {Phys. Rev. Lett.}\ }\textbf {\bibinfo
  {volume} {110}},\ \bibinfo {pages} {127205} (\bibinfo {year}
  {2013})}\BibitemShut {NoStop}%
\bibitem [{\citenamefont {Gong}\ \emph {et~al.}(2013)\citenamefont {Gong},
  \citenamefont {Sheng}, \citenamefont {Motrunich},\ and\ \citenamefont
  {Fisher}}]{Gong2013}%
  \BibitemOpen
  \bibfield  {author} {\bibinfo {author} {\bibfnamefont {S.-S.}\ \bibnamefont
  {Gong}}, \bibinfo {author} {\bibfnamefont {D.~N.}\ \bibnamefont {Sheng}},
  \bibinfo {author} {\bibfnamefont {O.~I.}\ \bibnamefont {Motrunich}},\ and\
  \bibinfo {author} {\bibfnamefont {M.~P.~A.}\ \bibnamefont {Fisher}},\
  }\bibfield  {title} {\bibinfo {title} {Phase diagram of the
  spin-$\frac{1}{2}$ ${J}_{1}$-${J}_{2}$ heisenberg model on a honeycomb
  lattice},\ }\href {https://doi.org/10.1103/PhysRevB.88.165138} {\bibfield
  {journal} {\bibinfo  {journal} {Phys. Rev. B}\ }\textbf {\bibinfo {volume}
  {88}},\ \bibinfo {pages} {165138} (\bibinfo {year} {2013})}\BibitemShut
  {NoStop}%
\bibitem [{\citenamefont {Castro}\ \emph {et~al.}(2006)\citenamefont {Castro},
  \citenamefont {Peres}, \citenamefont {Beach},\ and\ \citenamefont
  {Sandvik}}]{Castro2006}%
  \BibitemOpen
  \bibfield  {author} {\bibinfo {author} {\bibfnamefont {E.~V.}\ \bibnamefont
  {Castro}}, \bibinfo {author} {\bibfnamefont {N.~M.~R.}\ \bibnamefont
  {Peres}}, \bibinfo {author} {\bibfnamefont {K.~S.~D.}\ \bibnamefont
  {Beach}},\ and\ \bibinfo {author} {\bibfnamefont {A.~W.}\ \bibnamefont
  {Sandvik}},\ }\bibfield  {title} {\bibinfo {title} {Site dilution of quantum
  spins in the honeycomb lattice},\ }\href
  {https://doi.org/10.1103/PhysRevB.73.054422} {\bibfield  {journal} {\bibinfo
  {journal} {Phys. Rev. B}\ }\textbf {\bibinfo {volume} {73}},\ \bibinfo
  {pages} {054422} (\bibinfo {year} {2006})}\BibitemShut {NoStop}%
\bibitem [{\citenamefont {Albuquerque}\ \emph {et~al.}(2011)\citenamefont
  {Albuquerque}, \citenamefont {Schwandt}, \citenamefont {Het\'enyi},
  \citenamefont {Capponi}, \citenamefont {Mambrini},\ and\ \citenamefont
  {L\"auchli}}]{Albuquerque2011}%
  \BibitemOpen
  \bibfield  {author} {\bibinfo {author} {\bibfnamefont {A.~F.}\ \bibnamefont
  {Albuquerque}}, \bibinfo {author} {\bibfnamefont {D.}~\bibnamefont
  {Schwandt}}, \bibinfo {author} {\bibfnamefont {B.}~\bibnamefont {Het\'enyi}},
  \bibinfo {author} {\bibfnamefont {S.}~\bibnamefont {Capponi}}, \bibinfo
  {author} {\bibfnamefont {M.}~\bibnamefont {Mambrini}},\ and\ \bibinfo
  {author} {\bibfnamefont {A.~M.}\ \bibnamefont {L\"auchli}},\ }\bibfield
  {title} {\bibinfo {title} {Phase diagram of a frustrated quantum
  antiferromagnet on the honeycomb lattice: Magnetic order versus valence-bond
  crystal formation},\ }\href {https://doi.org/10.1103/PhysRevB.84.024406}
  {\bibfield  {journal} {\bibinfo  {journal} {Phys. Rev. B}\ }\textbf {\bibinfo
  {volume} {84}},\ \bibinfo {pages} {024406} (\bibinfo {year}
  {2011})}\BibitemShut {NoStop}%
\bibitem [{\citenamefont {Hamer}\ \emph {et~al.}(1991)\citenamefont {Hamer},
  \citenamefont {Oitmaa},\ and\ \citenamefont {Weihong}}]{Hamer1991}%
  \BibitemOpen
  \bibfield  {author} {\bibinfo {author} {\bibfnamefont {C.~J.}\ \bibnamefont
  {Hamer}}, \bibinfo {author} {\bibfnamefont {J.}~\bibnamefont {Oitmaa}},\ and\
  \bibinfo {author} {\bibfnamefont {Z.}~\bibnamefont {Weihong}},\ }\bibfield
  {title} {\bibinfo {title} {Zero-temperature properties of the quantum xy
  model with anisotropy},\ }\href {https://doi.org/10.1103/PhysRevB.43.10789}
  {\bibfield  {journal} {\bibinfo  {journal} {Phys. Rev. B}\ }\textbf {\bibinfo
  {volume} {43}},\ \bibinfo {pages} {10789} (\bibinfo {year}
  {1991})}\BibitemShut {NoStop}%
\end{thebibliography}%

\end{document}